\documentclass[fleqn,usenatbib]{mnras}

\usepackage[T1]{fontenc}
\usepackage{physics}
\usepackage{xcolor}
\usepackage{soul}
\usepackage{comment}

\DeclareRobustCommand{\VAN}[3]{#2}
\let\VANthebibliography\thebibliography
\def\thebibliography{\DeclareRobustCommand{\VAN}[3]{##3}\VANthebibliography}

\usepackage{graphicx}	
\usepackage{amsmath}	
\usepackage{amssymb}	
\usepackage{siunitx}
\usepackage{newtxtext,newtxmath}

\newcommand{\SNR}{\mathrm{SNR}}
\newcommand{\hmpc}{~h^{-1}\mathrm{Mpc}}
\newcommand{\reffig}[1]{Fig.~\ref{#1}}
\newcommand{\refeq}[1]{Eq.~\eqref{#1}}

\newcommand{\refsec}[1]{Section~\ref{#1}}
\newcommand{\patchy}{\textsc{Patchy}}
\DeclareMathOperator*{\argmax}{argmax}

\title[Cosmic Void Selection Effects]{Cosmic Void Baryon Acoustic Oscillation Measurement: Evaluation of Sensitivity to Selection Effects}

\author[D. Forero-S\'anchez et al.]{
Daniel Forero-S\'anchez,$^{1}$\thanks{E-mail: daniel.forerosanchez@epfl.ch}
Cheng Zhao,$^{1}$\thanks{E-mail: cheng.zhao@epfl.ch} Charling Tao,$^{2}$ Chia-Hsun Chuang,$^{3}$ Francisco-Shu Kitaura,$^{4,5}$
\newauthor Andrei Variu,$^{1}$ Amélie Tamone,$^{1}$ Jean-Paul Kneib$^{1,6}$
\\
$^{1}$Institute of Physics, Laboratory of Astrophysics, Ecole Polytechnique F\'ed\'erale de Lausanne (EPFL), Observatoire de Sauverny, CH-1290 Versoix, Switzerland\\
$^{2}$Aix Marseille Univ, CNRS/IN2P3, CPPM, 163, avenue de Luminy - Case 902 - 13288 Marseille Cedex 09, France\\
$^{3}$Kavli Institute for Particle Astrophysics and Cosmology, Stanford University, 452 Lomita Mall, Stanford, CA 94305, USA\\
$^{4}$Instituto de Astrof\'isica de Canarias, s/n, E-38205, La  Laguna, Tenerife, Spain\\
$^{5}$Departamento  de  Astrof\'isica, Universidad de La Laguna,  E-38206, La Laguna, Tenerife, Spain\\
$^{6}$Aix Marseille Universit\'e, CNRS, LAM (Laboratoire d'Astrophysique de Marseille) UMR 7326, 13388, Marseille, France
}

\date{Accepted XXX. Received YYY; in original form ZZZ}

\pubyear{2021}

\begin{document}
\label{firstpage}
\pagerange{\pageref{firstpage}--\pageref{lastpage}}
\maketitle

\begin{abstract}
Cosmic voids defined as a subset of Delaunay Triangulation (DT) circumspheres have been used to measure the Baryon Acoustic Oscillations (BAO) scale; providing tighter constraints on cosmological parameters when combined with matter tracers. These voids are defined as spheres larger than a given radius threshold, which is constant over the survey volume. However, the response of these void tracers to observational systematics has not yet been studied. In this work we analyse the response of void clustering to selection effects. We find for the case of moderate (<20 per cent) incompleteness, void selection based on a constant radius cut yields robust measurements. This is particularly true for BAO-reconstructed galaxy samples, where large-scale void exclusion effects are mitigated. Moreover, we observe for the case of severe (up to 90 per cent) incompleteness -- such as can be found at the edges of the radial selection function -- that an accurate estimation of the void distribution is necessary for unbiased clustering measurements. In addition, we find that without reconstruction, using a constant threshold under these conditions produces a stronger void exclusion effect that can affect the clustering on large scales. A new void selection criteria dependent on the (local) observed tracer density that maximises the BAO peak significance prevents the aforementioned exclusion features from contaminating the BAO signal. Finally, we verify, with large simulations including light cone evolution, that both void sample definitions (local and constant) yield unbiased and consistent BAO scale measurements.
\end{abstract}




\section{Introduction}
Baryon Acoustic Oscillations (BAO) are a signal encoded in the two-point statistics of the CMB temperature and the distribution of matter in the Universe. They provide a standard ruler \citep{Blake2003ProbingRuler, Seo2005BaryonicSurveys} which can be used to trace the evolution of our Universe back to the recombination epoch, about $\num{380000}~\mathrm{yr}$ after Big-Bang. This BAO signal was first observed on galaxy distributions by \citet{Eisenstein2005} on Sloan Digital Sky Survey (SDSS) data by estimating the two-point correlation function (2PCF) of the galaxy sample. Similarly \citet{Cole2005TheImplications} observed the signal by analysing the power spectrum of the 2dF survey galaxy sample. Since then, more observations have improved or complemented the early measurement such as the Baryon Oscillation Sky Survey (BOSS) \citep{Dawson2013} and the extended BOSS (eBOSS) \citep{Dawson2016}. The ongoing Dark Energy Spectroscopic Instrument (DESI) survey \citep{Levi2019TheDESI},  also aims at the measurement of this scale and will soon provide larger amounts of data than ever before. 

A variety of tracers can be used to estimate the matter density observationally. Their particular physical properties induce a bias in their clustering with respect to the underlying dark matter distribution. In addition, \citet{Gil-Marin2010ReducingStochasticity} have shown that a multi tracer approach can yield more information than a single tracer analysis. In particular, cosmic voids can be regarded as a secondary tracer that can be extracted directly from an existing tracer sample without further observations. 

In general, we define cosmic voids as underdensities in the matter field. In practice, however, the formal definition depends strongly on how voids are extracted from galaxy catalogues. Some methods rely on a continuous density field. They extract voids of arbitrary shape by merging underdense regions of the smoothed field (e.g. \citet{Platen2007}) or define strictly spherical voids (e.g. \citet{Colberg2005}). There are also methods using purely geometrical arguments to define voids from a discrete point distribution. For example, the \textsc{Zobov} \citep{Neyrinck2008} void finder, defines voids using the Voronoi tessellation of the point set. In the present work, we use the Delaunay Triangulation Void findEr (\textsc{Dive}) \citep{Zhao2016} which defines voids as a subset of the circumspheres of the simplices resulting from the Delaunay Triangulation \citep[DT;][]{Delaunay1934} of the point space. This definition in particular, ensures that there are no galaxies within the spheres and allows voids to overlap with each other.

\citet{Zhao2016} identify two distinct sphere populations that can be distinguished by their radius $R$. Spheres smaller than a threshold radius ($R_c$) behave as voids-in-clouds \citep{Sheth2003} since they are predominantly located in matter-dense regions. Hence, they do not fit the `underdensities in the field' definition of a cosmic void. Large spheres ($R>R_c$) are located in underdense regions and behave as voids-in-voids. These are the cosmic voids with which \citet{Kitaura2016} and \citet{Liang2016} detected the BAO signal from cosmic voids for the first time. Through this work we will also consider the subsample of spheres with $R>R_c$ as cosmic voids. The $R_c$ quantity is obtained through the optimisation of the BAO signal (see \refsec{sec:gen-methods}).

The allowance of overlapping spheres greatly increases the `void' sample size and removing the overlapping voids has been shown to eliminate the BAO signal \citep{Kitaura2016}.  Moreover, it has been observed that the void number density depends on the galaxy density \citep{Zhao2016} which implies that the sphere population and consequently the threshold $R_c$ will vary for tracers of different densities.

It has been shown that the void distribution encodes higher-order information \citep{Kitaura2016, Zhao2016} of the galaxy distribution. A multi-tracer analysis combining galaxies and voids \citep{Zhao2020} has shown that the void statistics can contribute to a decrease in the BAO scale measurement error in most mock realisations. In addition, there is a $\sim 10~{\rm per cent}$ improvement in the distance measurement for observational data obtained by the Sloan Digital Sky Survey (SDSS) corresponding to the BOSS DR12 LOW-Z sample. Furthermore, \citet{Zhao2021} obtain a 5 to 15~{\rm per cent} improvement in the BAO scale measurement on combined BOSS and eBOSS data.

The observed samples from redshift surveys are usually incomplete with respect to the number of known targets, due to effects like angular systematics, fiber collisions and redshift failures. These issues affect the sample on the angular plane and result in a relatively small incompleteness. Nonetheless, even small angular incompleteness may result in biased clustering measurements if not carefully corrected, as can be observed in galaxy clustering \citep{Ross2011, Ross2017a}. Besides these angular effects, there is further incompleteness that comes from the radial selection function (see for example fig. 11 in \citet{Reid2016}) which is accounted for in the analysis by means of the Feldman, Kaiser \& Peacock (FKP) weights \citep{Feldman1994}. For a complete discussion on the systematics of the BOSS DR12 sample see \citet{Reid2016}.

In this paper we aim to study in detail how the incompleteness in the tracer sample induced by observational systematics and radial selection affects the DT void sample and its clustering measurements. Additionally, we estimate how these may affect cosmological measurements involving voids.

In the next section, we describe the data used in this work. In \refsec{sec:gen-methods} we introduce the methods we used for the analysis such as the definition of the BAO Signal-to-Noise ratio (SNR), 2PCF estimator, definitions for the combined analysis, BAO reconstruction and BAO fitting. \refsec{sec:radius-cut-dependence} explores the dependence of the $\SNR$-maximising void radius cut on the tracer density. Later, we study how non-uniform incompleteness affects void clustering measurements (\refsec{sec:clustering-incompleteness}). Following this, in \refsec{sec:bao-constraints} we extend this analysis to BAO scale measurements. Finally, we conclude in \refsec{sec:conclusions} and outline future work.
\section{Data}
\label{sec:data}
In this section we introduce the galaxy catalogues used throughout this study. Additionally, we describe the downsampling techniques used to simulate radial selection and angular systematics on cubic mocks. Void catalogues are always obtained using the \textsc{DIVE}\footnote{\url{https://github.com/cheng-zhao/DIVE}} code or equivalently, its Python version \textsc{PyDIVE}\footnote{\url{https://github.com/dforero0896/pydive}}.

\subsection{Simulation boxes}
We use 500 realisations of mocks produced using the PerturbAtion Theory Catalogue generator of Halo and galaxY distributions (\patchy{}) code \citep{Kitaura2014}. The catalogues are produced using the Augmented Lagrangian Perturbation Theory and are calibrated using the BigMultiDark (MD) N-body simulations \citet{Klypin2016}. These are the snapshots used for generating BOSS DR12 MD-\patchy{} mocks. We will refer to this data as \textsc{Patchy} boxes. The boxes have a volume of $L^3=\num{2500}^3~\hmpc^3$ and have a galaxy density of  $\bar{n}_{\mathrm{gal,0}}\approx \num{4e-4}~h^3\mathrm{Mpc}^{-3}$.

The fiducial cosmology used for the construction of the mocks is \citep{Kitaura2016}
\begin{align}
    \label{eq:fiducial-om}
    &\Omega_{\rm m} = \num{0.307115},\\
    &\Omega_{\rm b} = \num{0.048206},\\
    &\sigma_8 = \num{0.8288},\\
    &n_{\rm s} = 0.9611,\\
    \label{eq:fiducial-h}
    &h = 0.677.
\end{align}
For \textsc{Patchy} boxes, we choose the radial direction to be along the $Z$ coordinate. We also generate \patchy{} boxes in redshift space by applying Redshift Space Distortions (RSD) in this direction. The angular plane corresponds then to $(X,\ Y)$.
\subsection{Simulating incompleteness in boxes}
Based on the \patchy{} boxes we create incomplete data via random downsampling of the complete point distribution. There are three downsampling schemes: uniform, angular and radial. The respective data sets are dubbed \patchy{}-U, \patchy{}-A and \patchy{}-R. To avoid confusion, we will refer to the full \patchy{} boxes (without downsampling) as \patchy{}-F.

In Table \ref{tab:labels} we summarise the galaxy data samples used and generated  for this study as well as the downsampling methods devised to generate them.
\begin{table}
    \centering
    \caption{Summary of samples used and generated for this work and the method used to generate them.}
    \begin{tabular}{lll}
    \hline
    \multicolumn{1}{c}{\textbf{Sample Label}} & \multicolumn{1}{c}{\textbf{Downsampling}} & \multicolumn{1}{c}{\textbf{Description}}                                      \\ \hline
    \patchy{}-F                                & None                                       & \begin{tabular}[c]{@{}l@{}}500 original\\ \textsc{Patchy} boxes\end{tabular}  \\ \hline
    \patchy{}-U                                & Uniform                                    & \begin{tabular}[c]{@{}l@{}}500 boxes\\ for each completeness $C$\\with uniform downsampling\end{tabular} \\ \hline
    \patchy{}-A                                & Angular (Parabolic)                        & \begin{tabular}[c]{@{}l@{}}500 boxes\\ with angular downsampling\end{tabular} \\ \hline
    \patchy{}-R                                & Radial (Gaussian)                          & \begin{tabular}[c]{@{}l@{}}500 boxes\\ with radial downsampling\end{tabular}  \\ \hline
    \end{tabular}
    \label{tab:labels}
\end{table}

\subsubsection{Uniformly downsampled boxes: \patchy{}-U}
The \patchy{}-U sample is created by downsampling the \patchy{} boxes with a constant completeness $C$. We vary $C$ in intervals of $0.05$ from $C=0.1$ to $C=0.95$, thus the resulting boxes have galaxy densities of $\bar{n}'_\mathrm{gal} = C \bar{n}_{\rm gal,0}$. The downsampled boxes have same clustering amplitude than the complete box but the variances increase. Under these conditions, we are able to study the dependence of the void sample purely on the tracer density.

\begin{figure}
    \centering
    \includegraphics[width=\linewidth]{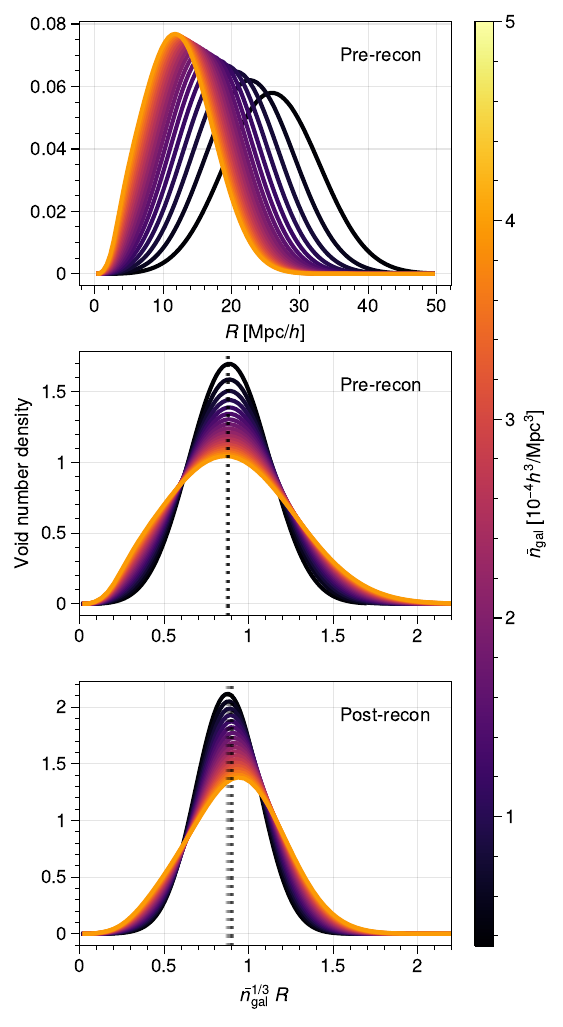}
    \caption{Void number distributions as a function of void radius for mocks of different uniform galaxy densities. The top panel shows the distributions as a function of the sphere radius $R$. The middle panel shows that in the dimensionless units $\bar{n}_{\mathrm{gal}}^{1/3}R$ the distributions span a similar range of dimensionless radii. The vertical dotted lines show the median void radius, the near-perfect alignment of the lines makes it difficult to distinguish them individually. Bottom panel shows the analogous plot to the mid panel for reconstructed boxes. Notice that in this case the mean values do not align as well as they do before reconstruction.}
    \label{fig:void-density-vs-radius}
\end{figure}

\reffig{fig:void-density-vs-radius} shows the void number density distributions as a function of the void radius (top panel) for \patchy{}-U catalogues with different densities. They shift towards larger voids as the galaxy density decreases, which is consistent with results shown in \citet{Zhao2016}. This suggests that the radius cut should depend on the galaxy density as well. This dependence can be parameterized as a power law, inspired by the definition of a dimensionless scale $\bar{n}_{\mathrm{gal}}^{1/3}R$. The bottom panel of the same figure shows that the distributions span a similar value range in this scale and the median void radii (black dotted lines) align at $R = \qty(0.8775\pm0.004)~\bar{n}_{\mathrm{gal}}^{-1/3}$.

\subsubsection{Non-Uniform Angular downsampling: \patchy{}-A}
\begin{figure*}
    \centering
    \includegraphics[width=\linewidth]{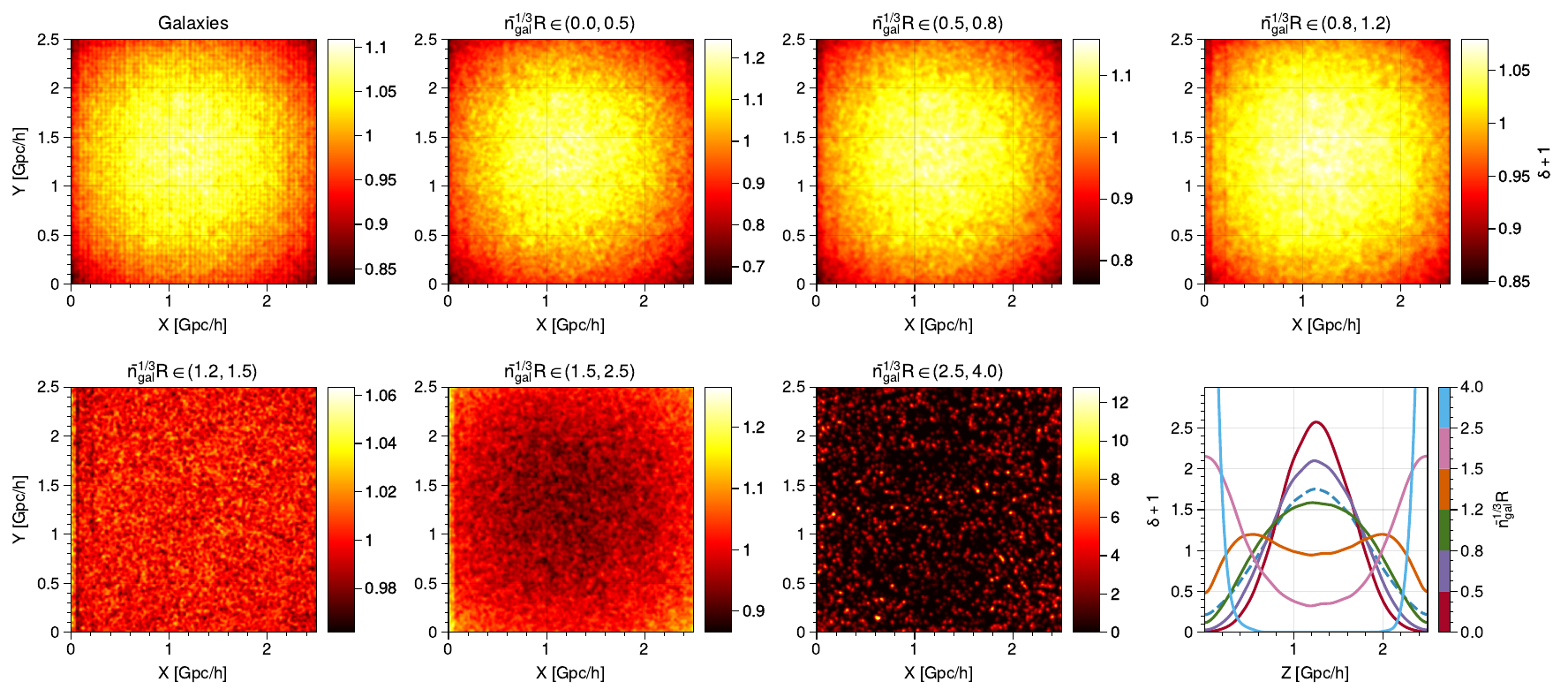}
    \caption{Tracer densities for boxes with simulated incompleteness. Color maps show the density field $\delta+1 = \rho/\bar{\rho}$ for galaxies (top left) with angular downsampling. The following maps show the density fields for DT spheres in different radius bins. The last panel shows the radial density ($Z$ direction) for galaxies downsampled in this direction (dashed line) and the DT spheres extracted from it in different radius bins (continuous lines). The colorbar shows the radius bin corresponding to each curve.}
    \label{fig:box-incompleteness}
\end{figure*}
Angular systematics on observations can be essentially described as sample incompleteness in the angular plane. Recent cosmological clustering analyses focus on regions of the sky with over $90~{\rm per cent}$ completeness \citep{Ross2020}, such that the sample incompleteness is very low. In this work we simulate this effect by downsampling the \patchy{} boxes in the $X,~Y-$plane with a 2D paraboloid filter. The parabolic map has a maximum completeness of 1 in the centre of the plane and minima of 0.8 in the corners. \reffig{fig:box-incompleteness} shows the densities of galaxies (top-left panel) as well as DT spheres with different radii. It should be noted that the distributions of smaller spheres ($R<1.2\bar{n}_{\rm gal}^{-1/3}$) are close to that of the galaxies of \patchy{}-A. As the sphere radius increases ($R\in(1.2, 1.5)\bar{n}_{\rm gal}^{-1/3}$) the distribution becomes very uniform (bottom-left panel) and for even larger radii ($R>1.5\bar{n}_{\rm gal}^{-1/3}$), it becomes anti correlated to that of the galaxies (bottom middle-left panel).
\subsubsection{Non-Uniform Radial downsampling: \patchy{}-R}
To simulate a realistic radial selection function, we use a Gaussian function to downsample in the radial direction:
\begin{equation}
\label{eq:radialincompleteness}
C(Z)=\frac{n'_{\mathrm{gal}}}{\bar{n}_{\mathrm{gal}}}(Z) = \exp{-\frac{(Z-c_ZL)^2}{2(\sigma_ZL)^2}}.
\end{equation}
The parameters $\quad c_Z=0.5,~\sigma_Z=0.235$ define respectively the centre and width of the distribution and $L$ is the box size. The value of $\sigma_Z$ is chosen in such a way that at the box boundaries, $\sim 10~{\rm per cent}$ of the galaxies are kept. The bottom-right panel of \reffig{fig:box-incompleteness} shows the measured radial densities from \patchy{}-R boxes (dashed line) and the corresponding void density for different radius bins (solid lines). Analogous to the angular maps, spheres with radius ranges of $R<1.2\bar{n}_{\rm gal}^{-1/3}$, $R\in(1.2, 1.5)\bar{n}_{\rm gal}^{-1/3}$ and $R>1.5\bar{n}_{\rm gal}^{-1/3}$ are correlated, independent of and anti-correlated with the galaxy field respectively.

\subsection{BOSS data and mocks}
Besides mock simulation boxes, we analyse light cone data. We use the BOSS DR12 Luminous Red Galaxy (LRG) sample and the corresponding mock catalogues, the \textsc{MD-Patchy DR12} data. The BOSS DR12 LRG sample is composed of 1.2 million objects spanning $\num{9300}~\rm deg^2$ and redshifts $z\in(0.2,0.75)$ \citep{Ross2017a}. It is the result of observations done from the Apache Point Observatory up to phase III of SDSS. The corresponding mock catalogues are created by calibrating the Halo Abundance Matching (HAM) to match Big MultiDark with BOSS DR12 LRG data \citep{Kitaura2016}. 

\begin{figure}
    \centering
    \includegraphics[width=\linewidth]{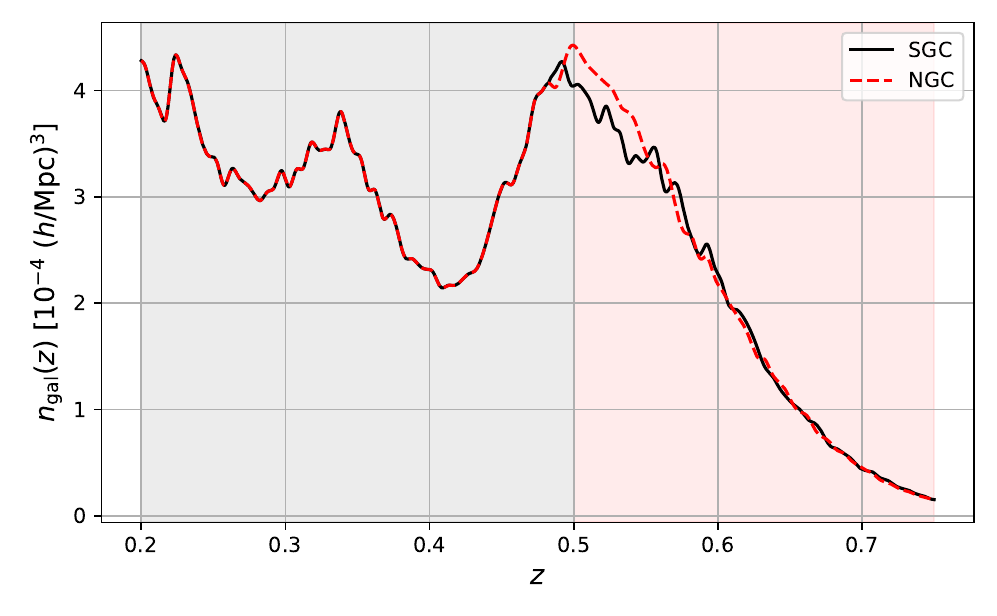}
    \caption{Radial number densities for BOSS DR12 data in the North (NGC) and South (SGC) galactic caps. Measured from \patchy{} mocks. The two redshift bins used in the analysis are shaded in gray (low-$z$) and red (high-$z$).}
    \label{fig:nz-caps}
\end{figure}

For our analysis, we use the combined (LOWZ and CMASS) sample and divide it in two redshift bins \citep{Alam2017TheSample, Zhao2020}, a low-$z$ bin with $z\in(0.2, 0.5)$ and a high-$z$ bin with $z\in(0.5, 0.75)$. To correct for known systematic effects in the galaxy sample, we also use weighting schemes as thoroughly described \citet{Reid2016, Ross2017a}. FKP weights \citep{Feldman1994} are also included to reduce the variance due to the radial selection function. \reffig{fig:nz-caps} shows the radial distribution of galaxies on light cones.

We use 2048 pre-reconstruction \textsc{MD-Patchy DR12} mocks and 1000 reconstructed light cones to estimate the error on the observations.

To perform BAO fits on light cones we use the fiducial cosmology $h=0.676$ and $\Omega_{\rm m}=0.31,\ \Omega_{\rm b} = \num{0.048},\ \sigma_8 = \num{0.8},\ n_{\rm s} = 0.97$.

\section{Methods}
\label{sec:gen-methods}

\subsection{BAO Reconstruction}

We apply BAO reconstruction \citep{Eisenstein2007a} to both cubic simulations and light cone data since it is the standard method for obtaining the tightest BAO constraints. We employ the iterative Fourier-space method described in \citet{Burden2015}. 

The idea behind reconstruction is to recover the initial positions $\vb*q$ of the particles in some fluid from their current (measured) positions $\vb*{x}$. This amounts to finding the displacement field $\vb*{\Psi}(\vb*{q})$ that relates them through $\vb*{x}(\vb*{q}) = \vb*{q} + \vb*{\Psi}(\vb*{q})$. 

The first-order solution to the displacement field is the Zeldovich approximation \citep{Zeldovich1970}. Given a real-space overdensity field $\delta_r(\vb*{x})$ one can obtain the displacement field by solving $\nabla\cdot \vb*{\Psi}(\vb*{q}) = -\delta_r(\vb*{x})/b$ in Fourier space to obtain
\begin{equation}
    \vb*{\Psi}(\vb*{k}) = i\frac{\vb*{k}}{k^2}\frac{\delta_r(\vb*{k})}{b},
    \label{eq:displacement-zeldovich}
\end{equation}
where $b$ is the linear galaxy bias, and $\vb*{\Psi(\vb*{k}})$ and $\delta_r(\vb*{k})$ are the displacement and real-space overdensity fields in Fourier space. 

To obtain the displacement field from a redshift-space overdensity, the process must account for RSD. The equation to solve to consider this effect is instead \citep{Nusser1993OnSamples}
\begin{equation} 
    \label{eq:displacement-rsd}
    \nabla\cdot\vb*{\Psi} + \frac{f}{b}\nabla\cdot(\vb*{\Psi}\cdot\vb*{\hat{r}})\vb*{\hat{r}} = -\frac{\delta_s(\vb*{x})}{b},
\end{equation}
with $f$ the growth rate, $\vb*{r}$ the radial coordinate vector, $\vb*{\hat{r}}$ the unitary vector of $\vb*{r}$ and $\delta_s$
is the redshift-space overdensity. The solution is obtained iteratively by using the following update rules:
\begin{enumerate}
    \item Estimate the real-space density field at iteration $i$, $\delta_r^{(i)}(\vb*{x})$, by applying a mass assignment scheme to the set of particles at positions $\{\vb*{x}^{(i)}\}$ and smoothing it.
    \item Estimate the displacement field using the Zeldovich approximation
    \begin{equation}
        \label{eq:psi-update}
        \vb*{\Psi}^{(i+1)}(\vb*{x}) = \mathcal{F}^{-1}\qty{i\frac{\vb*{k}}{k^2}\frac{\delta_r^{(i)}(\vb*{k})}{b}},
    \end{equation}
    where $\mathcal{F}^{-1}$ denotes the inverse Fourier transform.
    \item Partially remove RSD by updating the particle positions through 
    \begin{equation}
        \label{eq:pos-update}
        \vb*{x}^{(i+1)}=\vb*{x}^{(i)} + f \vb*{\hat{r}}^{(i)}\cdot\vb*{\Psi}^{(i+1)}(\vb*{x}^{(i)}).
    \end{equation}
\end{enumerate}
The initial condition of the iteration in given by $\vb*{x}^{(0)}\equiv\vb*{x}_s$, the redshift-space positions. After $n$ iterations, the density field will closely approximate the real-space density field, such that the displacement field computed from it using \refeq{eq:displacement-zeldovich} is a good approximation to the real displacement field. \citet{Burden2015} show that $n=3$ provides a good approximation.

In this work we use the \textsc{Revolver}\footnote{\url{https://github.com/seshnadathur/Revolver}} \citep{Nadathur2019REVOLVER:Reconstruction} implementation of this algorithm with $n=3$ iterations. The linear bias of mock samples is $b=2.2$. The effective redshifts of the simulation boxes and light cones are 0.466 and 0.5 respectively, which correspond to growth rate values of $f=0.743$ and $f=0.757$. The overdensity field is put in a grid of size $512^3$ using the Cloud-In-Cell (CIC) mass assignment scheme. A smoothing scale of $15\hmpc$ is used.

\subsection{2PCF estimation}
\label{sec:2pcf-estimation}
We use two different estimators for the 2PCF $\xi^{\rm ab}(s)$ between tracers $\rm a$ and $\rm b$ as a function of separation $s$. When we assume uniform random catalogues in a periodic box, we use the natural estimator \citep{Peebles1974}; when a random catalogue is needed to account for geometric effects or inhomogeneity of the galaxy density, we use the Landy-Szalay estimator \citep{Landy1993, Szapudi1997ACorrelations, Padmanabhan2012}. 

In practice, the random catalogues are built such that they account for all known systematic effects or sources of inhomogeneity in the sample. However, unknown systematics would never be included in the randoms. We simulate the presence of unknown systematics by using uniform randoms with inhomogeneous data. For example, computing the 2PCF of the \patchy{}-A data with uniform randoms would simulate the effect of an unknown source of moderate sample incompleteness.

The natural estimator is defined as follows:
\begin{equation}
    \xi^{\rm ab}(s) = \frac{{\rm D_a}{\rm D_b}(s)}{\rm{RR}(s)} - 1,
\end{equation}
where ${\rm RR}(s)$ indicates the normalised random-random pair counts which can be computed analytically. The ${\rm D_a}{\rm D_b}(s)$ are the normalised pair counts between data catalogues of tracers a and b.

On the other hand, the Landy-Szalay (LS) estimator is defined as
\begin{equation}
    \label{eq:ls-estimator}
    \xi^{\rm ab}(s) =\rm  \frac{D_{\rm a}D_{\rm b} - D_{\rm a}R_{\rm b} - D_{\rm b}R_{\rm a} + R_{\rm a}R_{\rm b}}{R_{\rm a}R_{\rm b}}.
\end{equation}
The quantity ${\rm R_a} {\rm R_b}$ represents the normalised random-random pair counts for tracers $\rm a$ and $\rm b$. Similarly, ${\rm D_a} {\rm R_b}$ are the data-random normalised pair counts between data catalogue of tracer a and random catalogue of tracer b. This estimator is also used for reconstructed catalogues, in which case we replace the galaxy random catalogues ${\rm R_g}$ with the shifted random ${\rm S_g}$ in the numerator.

For the light cone catalogues we use a modified version of the estimator introduced by \citet{Zhao2020} for the combined 2PCF of two tracers a and b. We use a pair of weights $w_{\rm a}$ and $w_{\rm b} = 1 - \abs{w_{\rm a}}$, such that the a-a and b-b correlations are recovered for $w_{\rm a} = -1$ and $w_{\rm a} = 0$ respectively. We combine each term in the LS estimator using the following formulae
\begin{equation}
\begin{split}
    {\rm D_aD_{b}^{comb}}& = \frac{w_{\rm a}^2N_{\rm D_aD_a}{\rm D_aD_a} + w_{\rm a}w_{\rm b}N_{\rm D_aD_b}{\rm D_aD_b} + w_{\rm b}^2N_{\rm D_bD_b}{\rm D_bD_b}}{w_{\rm a}^2N_{\rm D_aD_a} + w_{\rm a}w_{\rm b}N_{\rm D_aD_b} + w_{\rm b}^2N_{\rm D_bD_b}},
\end{split}
\end{equation}
\begin{equation}
\begin{split}
    {\rm D_aR_{b}^{comb}}& = \frac{w_{\rm a}^2N_{\rm D_aR_a}{\rm D_aR_a} + w_{\rm a}w_{\rm b}N_{\rm D_aR_b}{\rm D_aR_b} + w_{\rm b}^2N_{\rm D_bR_b}{\rm D_bR_b}}{w_{\rm a}^2N_{\rm D_aR_a} + w_{\rm a}w_{\rm b}N_{\rm D_aR_b} + w_{\rm b}^2N_{\rm D_bR_b}},
\end{split}
\end{equation}
\begin{equation}
\begin{split}
    {\rm R_aR_{b}^{comb}}& = \frac{w_{\rm a}^2N_{\rm R_aR_a}{\rm R_aR_a} + w_{\rm a}w_{\rm b}N_{\rm R_aR_b}{\rm R_aR_b} + w_{\rm b}^2N_{\rm R_bR_b}{\rm R_bR_b}}{w_{\rm a}^2N_{\rm R_aR_a} + w_{\rm a}w_{\rm b}N_{\rm R_aR_b} + w_{\rm b}^2N_{\rm R_bR_b}}.
\end{split}
\end{equation}
and combined correlation is then given by the LS estimator in \refeq{eq:ls-estimator}.

We have omitted the explicit dependence on the separation $s$ for brevity. The quantities $N_{\rm AB}$ denote the normalisation of the pair counts $\rm AB$, which are given by $N_{\rm AA}\equiv n_{\rm A}(n_{\rm A}-1)$ and $N_{\rm AB}\equiv n_{\rm A}n_{\rm B}$. Where $n_{\rm A}$ denotes the weighted number of objects in catalogue A. In particular, we combine correlations of galaxies and voids by identifying tracer a with voids and tracer b with galaxies. 

To generate random catalogues for voids ($\rm R_{\rm v}$) we apply a slightly modified version of the `shuffling' algorithm presented in \citet{Liang2016}. It aims to reproduce the radial selection function and angular incompleteness for voids of different radii by assigning a redshift, radius pair to a random void angular position. In our implementation for the mocks with radial downsampling, we assign $Z$, radius and $n_{{\rm gal}}(Z)$ values to a random $X,\ Y$ angular void position. For the \patchy{}-A mocks we assign $X$, $Y$, radius and $n_{\rm gal}(X,Y)$ values to random $Z$ void positions.

As a quick and model-independent estimate of the `goodness' of the BAO significance (see Appendix~\ref{sec:snr-def} for more details), we use the BAO signal definition introduced by \citet{Liang2016}:
\begin{equation}
S = \xi(s^{\mathrm{BAO}}) - \frac{\xi(s_1^{\mathrm{dl}}) + \xi(s_2^{\mathrm{dl}}) + \xi(s_1^{\mathrm{dr}}) +  \xi(s_2^{\mathrm{dr}})}{4},
\end{equation}
with $\xi(s)$ the correlation function. $s^{\mathrm{BAO}}$, $s_1^{\mathrm{dl}}$, $s_2^{\mathrm{dl}}$, $s_1^{\mathrm{dr}}$ indicate the abscissas near the BAO peak and the associated left and right dips of the 2PCFs of mocks. Their values are chosen to be $102.5$, $82.5$, $87.5$, $117.5$ and $122.5\hmpc$ respectively for 2PCFs measured in 40 separation bins from 0 to $200\hmpc$. Moreover, we average two bins for both the left and right dips to increase the robustness in the measurement of $S$.

With this definition, we can estimate the $\SNR$ of the BAO peak as 
\begin{equation}
\mathrm{SNR} = \frac{\langle S \rangle}{\sigma_S},
\label{eq:snr-estimate}
\end{equation}
where $\langle S \rangle$ and $\sigma_S$ denote the mean and standard deviation of the BAO signal computed over an ensemble of mock realisations. We further estimate the uncertainties of the SNR measurements using the jackknife method. This is done by leaving out one mock realisation for the SNR evaluation at one time. When the 2PCFs correspond to anti-correlated samples, both $S$ and $\SNR$ are negative.

Our $\SNR$ definition provides a robust and quick estimation of the goodness of the BAO signal. The results are generally consistent with the uncertainties of the BAO peak position inferred from BAO fits. For a further discussion about the definition of the $\SNR$ used here see Appendix~\ref{sec:snr-def} and \citet{Liang2016}.

\subsection{BAO fitting}
\subsubsection{BAO model}
\label{sec:baomodel}
To fit the BAO feature we use the model proposed by \citet{Xu2012}, which relies on a template correlation function defined as
\begin{equation}
\xi_{\rm t}(s) = \int\frac{\dd k}{2\pi^2}\frac{\sin ks}{ks}k^2P_{\rm t}(k)\exp(-k^2a^2),
\end{equation}
where the parameter $a = 1\hmpc$ \citep{Xu2012, Zhao2020} defines the exponential damping factor for ensuring the convergence of numerical integration. The fiducial cosmological information is encoded in the template power spectrum $P_{\rm t}(k)$:
\begin{equation}
P_{\rm t}(k) = \qty[P_{\mathrm{lin}}(k) - P_{\rm lin, nw}(k)]\exp\qty(\frac{-\Sigma_{\mathrm{nl}}^2k^2}{2}) + P_{\rm lin, nw}(k).
\label{eq:pkt}
\end{equation}
The functions $P_{\mathrm{lin}}(k)$ and $P_{\rm lin, nw}(k)$ denote the linear matter power spectrum and its non-wiggle (BAO free) counterpart. The latter is obtained from the fitting formulae provided by \citet{Eisenstein1997} given the fiducial cosmological parameters. Meanwhile, the linear power spectrum is computed using CAMB \citep{Lewis2002CosmologicalApproach}. For the cubic mocks, it is essentially the input power spectrum used to create the simulations. The $\Sigma_{\rm nl}$ quantity parameterizes the damping of the BAO signal due to non-linear effects.

The model correlation function that is fitted to the data is defined as \citep{Xu2012},
\begin{equation}
\xi_{\mathrm{model}}(s) \equiv B^2 \xi_{\rm t}(\alpha s) + A(s).
\label{eq:model2pcf}
\end{equation}
The $B$ parameter regulates the amplitude of the correlation function and $\alpha$ -- the dilation parameter --  is a measure of the position of the BAO peak relative to the theoretical prediction in the fiducial cosmology. Finally, the polynomial term
\begin{equation}
A(s) = \frac{a_1}{s^2} + \frac{a_2}{s} + a_3,
\end{equation}
includes three nuisance parameters $a_i,\ (i=1,2,3)$ that fit the broad shape of the 2PCF.

\citet{Zhao2020} have shown that this model does not fit void 2PCFs well. Consequently, they propose a generalised template power spectrum that takes into account the broad-band behaviour of the void non-wiggle power spectrum $P_{\rm v, nw}(k)$:
\begin{equation}
\begin{split}
    P_{\rm t}(k) =& \qty{\qty[P_{\mathrm{lin}}(k) - P_{\mathrm{lin,nw}}(k)]\exp\qty(\frac{-\Sigma_{\mathrm{nl}}^2k^2}{2}) + P_{\mathrm{lin,nw}}(k)}\times\\&\frac{P_{\mathrm{v,nw}}(k)}{P_{\mathrm{lin,nw}}(k)}.
\end{split}
\label{eq:pktmod}
\end{equation}
The function $P_{\rm v, nw}(k)$ is not trivial to compute and there is no analytic form for it yet. In this work, we follow \citet{Zhao2020} and parameterize it as
\begin{equation}
\label{eq:pktemplate}
\frac{P_{\mathrm{v,nw}}(k)}{P_{\mathrm{lin,nw}}(k)}=1+ck^2,
\end{equation}
where $c$ is an extra free parameter. This has been shown to be a good approximation on large scales. For a complete discussion on the modified BAO model see \citet{Zhao2020}.

Through this work we use the modified BAO model to fit the void auto correlations and galaxy-void combined correlation functions (see \refsec{sec:2pcf-estimation}). However, with this model it is observed that the presence of the $c$ parameter causes the $\alpha$ parameter to converge towards the edges of the prior. We avoid this by fitting first the average of the mocks in order to estimate $c$ accurately. We then fix the $c$ parameter to the best-fit estimate value when fitting individual mocks.

\subsubsection{Parameter inference}
The model described in \refsec{sec:baomodel} consists -- in principle -- of 7 free parameters for void and combined correlation functions. These are $\alpha,\ B,\ \Sigma_{\rm nl},\ c,\ a_1,\ a_2$ and $a_3$. The nuisance parameters are fitted using the least-squares method. 
For the remaining parameters $\Theta = \qty{\alpha,\ B,\ \Sigma_{\rm nl},\ c}$, we infer posterior distributions $p(\Theta|X)$ in the Bayesian approach:
\begin{equation}
p(\Theta|X) = \frac{p(X|\Theta)p(\Theta)}{p(X)},
\label{eq:bayes}
\end{equation}
where $X$ denotes the data. The distribution $p(X|\Theta)\equiv\mathcal{L}$ is the likelihood
\begin{equation}
\mathcal{L}(\Theta) \equiv A\exp(-\frac{\chi^2(\Theta)}{2}),
\end{equation}
where $A$ is a normalisation constant and $\chi^2(\Theta)$ is defined as
\begin{equation}
\chi^2(\Theta) = \qty[\xi_{\mathrm{data}} - \xi_{\mathrm{model}}(\Theta)]^T\vb*{C}^{-1}\qty[\xi_{\mathrm{data}} - \xi_{\mathrm{model}}(\Theta)].
\end{equation}
Here, $\xi_{\mathrm{data}}$ is the measured correlation function and $\xi_{\mathrm{model}}$ is the model 2PCF in \refeq{eq:model2pcf}. The matrix $\vb*{C}^{-1}$ is the inverse unbiased covariance matrix estimated from the mocks \citep{Hartlap2007}
\begin{equation}
\vb*{C}^{-1} = \vb*{C}_s^{-1}\frac{N_m - N_{\mathrm{bins}} - 2}{N_m - 1}.
\label{eq:covmat}
\end{equation}
The matrix $\vb*{C}_s^{-1}$ is the sample covariance matrix, 
\begin{equation}
\vb*{C}_{s,\ ij} =  \frac{1}{N_m -1}\sum_{k=1}^{N_m}[\xi_k(s_i) - \bar{\xi}(s_i)][\xi_k(s_j) - \bar{\xi}(s_j)],
\end{equation}
where $N_m$ is the number of mocks used, $N_{\mathrm{bins}}$ is the number of separation bins used for the fit, $\xi_k(s)$ is the 2PCF of the $k$-th mock catalogue, $\bar{\xi}(s)$ is the average 2PCF over all mocks and the indices $i,\ j = 1,\ \dots,\ N_{\mathrm{bins}}$. When using the modified BAO model to fit the average correlation function of the mocks we rescale the covariance by a factor of $N_m^{-1}$ in order to better constrain $c$.

The evidence $p(X)\equiv\mathcal{Z}(X)$ can be computed as the normalisation of the posterior
\begin{equation}
p(X)\equiv\mathcal{Z} = \int\dd\Theta p(\Theta)p(X|\Theta).
\end{equation}
It corresponds to the average likelihood over the prior distribution, $p(\Theta)$.

The \textsc{Multinest} algorithm \citep{Feroz2008} based on Nested Sampling \citep{Skilling2004} is used to compute the evidence and the posteriors in this work. We also use the \textsc{Baoflit}\footnote{\url{https://github.com/cheng-zhao/BAOflit}} code for fast integration and likelihood evaluations. The reported values of the parameters are the maximum likelihood estimates, and the errors correspond to the 1$\sigma$ estimate from the marginalised posteriors, which is obtained from the \texttt{pymultinest}'s analyser \citep{Buchner2014}. 

In our fits, we use the fitting range $s\in[60, 150]~\hmpc$ with $\num{1000}$ live points and tolerance of $\num{0.1}$ following the settings in \citet{Zhao2020}.

\subsubsection{Priors}
\label{sec:priors}
The prior distribution $p(\Theta)$ of parameter $\Theta$ depends entirely on our knowledge.

Following \citet{Zhao2020}, we set the following priors
\begin{align}
&p(\Sigma_{\mathrm{nl}}) = \mathcal{U}(0, 25)\\
&p(B) = \mathcal{U}(0, 50)\\
&p(\alpha)=\mathcal{U}(0.8, 1.2)
\end{align}

For void correlations from boxes we set the prior of the $c$ parameter to
\begin{align}
&p(c) = \mathcal{U}(0,1000),
\end{align}
and for combined correlations in light cones we set 
\begin{align}
&p(c) = \mathcal{U}(0,2000).
\end{align}
Here, $\mathcal{U}(\Theta_{\mathrm{min}}, \Theta_{\mathrm{max}})$ denotes a uniform distribution between $\Theta_{\mathrm{min}}$ and $\Theta_{\mathrm{max}}$. 

\section{Radius cut dependence on galaxy density}
\label{sec:radius-cut-dependence}
In this section we use the pre- and post-reconstruction \patchy{}-U redshift space data to infer the relation between the radius cut $R_c$, the galaxy density and the $\SNR$. We first establish the procedure to obtain $\SNR$ maps and maximise them to define an optimal radius cut $R_c^*$. Then, we apply this procedure to pre-reconstruction correlations to obtain $R_c^*(\bar{n}_{\mathrm{gal}})$. Finally we repeat the procedure with post-reconstruction large-sphere auto correlations and obtain an analogous fitting formula.

Through this section we often quote sphere radius values in terms of their dimensionless value $\bar{n}_{\mathrm{gal}}^{1/3}R$, when relevant, we also display the corresponding dimensionful value for a BOSS-compatible density of $\bar{n}_{\mathrm{gal}} \approx \num{3.5e-4}~h^3\mathrm{Mpc}^{-3}$.

\subsection{Maximisation of the $\abs{\SNR}$}
In order to construct the $\SNR(R_c, \bar{n}_{\mathrm{gal}})$ map, we compute the SNR from 500 correlation functions for each $(R_c, \bar{n}_{\mathrm{gal}})$ pair (see \refsec{sec:gen-methods}). For each density bin we examine $\bar{n}_{\mathrm{gal}}^{1/3}R_c$ values in the range of $[0.5, 1.33]$ for large spheres (radii $R > R_c$) and $[0.3, 0.9]$ for small spheres ($R < R_c$). The density bins correspond to the different completeness values used to create the \patchy{}-U data set (see \refsec{sec:data}).

Then, from the $\SNR$ maps, we define the optimal radius cut as
\begin{equation}
R_c^*(\bar{n}_{\mathrm{gal}}) = \argmax_{R_c}\left(\abs{\SNR(R_c, \bar{n}_{\mathrm{gal}})}\right).
\end{equation}
The maximisation is performed separately for each $\bar{n}_{\mathrm{gal}}$ by fitting a parabolic model to the points closest to the maximum $\abs{\SNR}$ using the \textsc{iminuit} package\footnote{\url{https://github.com/scikit-hep/iminuit}}. This yields a set of $(R_c^*, \bar{n}_{\mathrm{gal}})$, to which we fit a power law of the form 
\begin{equation}
    \label{eq:model-radii}
    R^*_c / (h^{-1}~{\rm Mpc})= \gamma (\bar{n}_{\mathrm{gal}}/(h~{\rm Mpc}^{-1}))^\nu,
\end{equation}
also using \textsc{iminuit}. The weight of each point in the fit is chosen to be to the measured $\abs{\SNR}$, which gives more importance to higher-density measurements which are more important for cosmological BAO studies. In what follows we assume the units of the radii and number densities are fixed and therefore omit them from equations hereafter.
\subsection{Pre-reconstruction}

\begin{figure*}
    \centering
    \includegraphics[width=\linewidth]{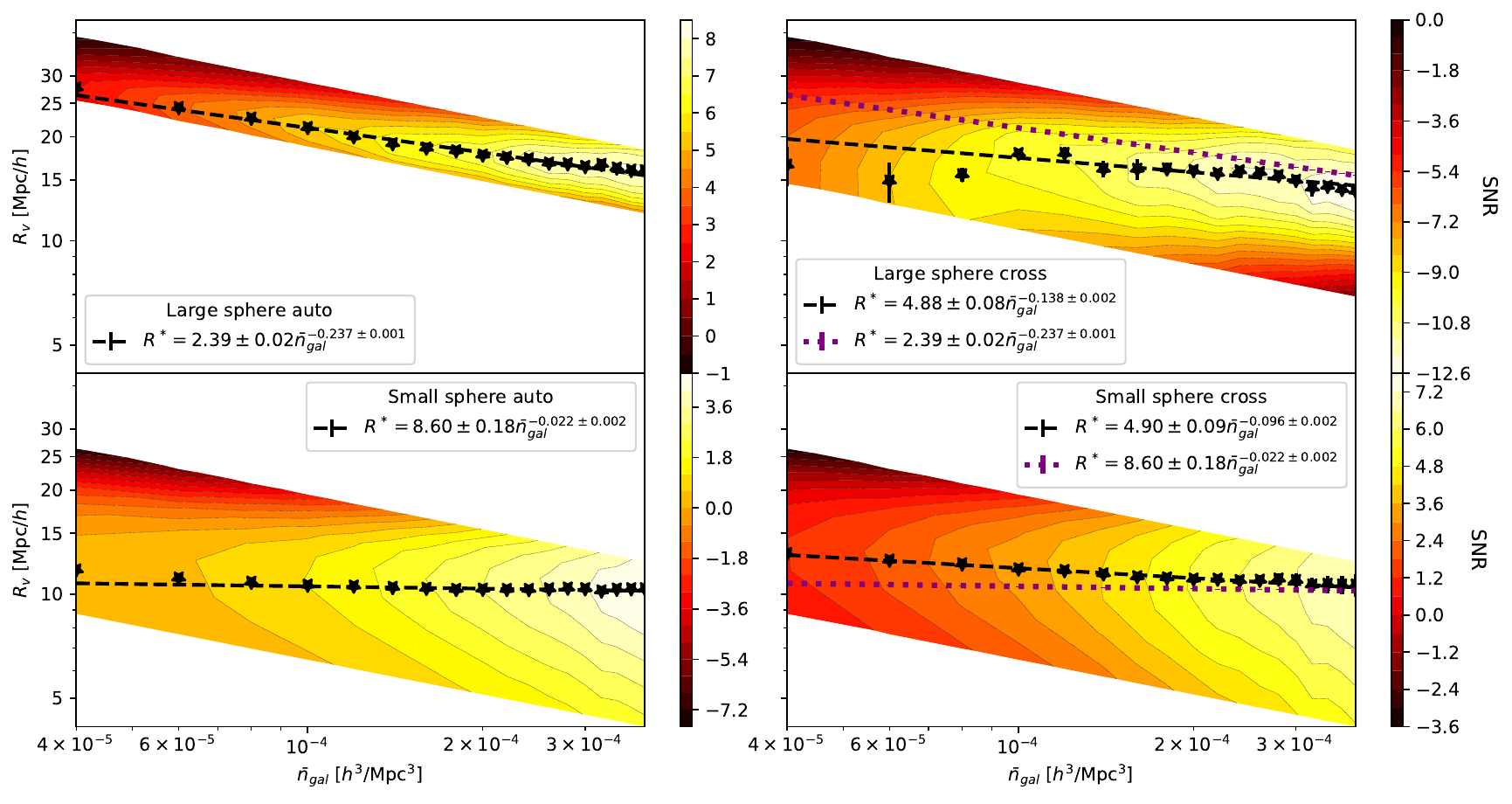}
    \caption{Optimum radius cut for pre-reconstruction boxes of different uniform densities. The contour map corresponds to the BAO $\SNR(R_c, \bar{n}_{\mathrm{gal}})$. The stars show the measured optimal cut $R_c^*$  as a function of the galaxy density and the dashed line shows the fit to the model $R^*_c = \gamma \bar{n}_{\mathrm{gal}}^\nu$. The top panels show results for large-spheres while the bottom panels show measurements for small-spheres. Left side panels correspond to auto correlations and right side panels to cross-correlations. The dotted lines on the cross correlation panels show the fit to the corresponding auto correlations. }
    \label{fig:snr-all-r-n-pre}
\end{figure*}
\subsubsection{Small-sphere auto correlation}
In \reffig{fig:snr-all-r-n-pre} (bottom-left panel) we present the results of the $\SNR$ optimisation for the small sphere sample. Our fit shows that for densities $\lesssim \num{e-4}~h^3~\mathrm{Mpc}^{-3}$ there is no BAO signal in this correlation function. For larger densities, the measured $\SNR \lesssim 4$, meaning that the small sphere auto correlation does not yield much information and may be difficult to fit. 

For small densities and large radius cuts we find a large negative $\SNR$, which shows that in this region there are negative correlations due to the inclusion of large-enough spheres which actually trace underdense regions. Finally we observe that the $R^*_c$ for this sample has a very weak dependence on the galaxy density and is given by
\begin{equation}
\begin{split}
    &R^*_c \approx (8.60\pm0.18)~~\bar{n}_{\mathrm{gal}}^{-0.022\pm0.002}\\
    &R_{c,\rm BOSS}^* = 10.25 \pm 0.27 \hmpc).    
\end{split}
\end{equation}

Due to the low $\SNR$ and the small dependence on the galaxy density, we do not consider this correlation further in this paper.

\subsubsection{Large-sphere auto correlation}
The results of the $\SNR$ map for void-void auto correlations are shown in the top-left panel of \reffig{fig:snr-all-r-n-pre}. Our model fit yields
\begin{equation}
    \begin{split}
        &R_c^* \approx (2.39\pm0.02)~\bar{n}_{\mathrm{gal}}^{-0.237\pm0.001}\\
        &R_{c,\rm BOSS}^* = 15.76 \pm 0.18 \hmpc.
    \end{split}
    \label{eq:pre-rdependence}
\end{equation}

The maximum $\SNR$ we obtain is $\sim 8$, smaller than the maximum galaxy-galaxy correlation $\SNR$, which is closer to $\sim 9.5$. 

Moreover, the corresponding value for BOSS-compatible density ($R_{c,\rm BOSS}^*$) is consistent with previous studies that suggest a radius cut of $15$ to $16 \hmpc$ \citep{Liang2016, Zhao2020}.

\subsubsection{Cross correlations between galaxies and small-spheres}
Cross correlating the small sphere sample with the galaxy population yields higher BAO $\SNR$ when compared to the sphere auto correlation. We find a maximum $\SNR_{\mathrm{max}} \sim 7$, as shown in the bottom-right panel of  \reffig{fig:snr-all-r-n-pre}, which is comparable to that of large sphere auto correlation.  Our fit results show that the optimum small sphere sample to cross correlate with galaxies is defined differently than that of the auto correlation fit. In this case we obtain that 
\begin{equation}
\begin{split}
    &R^*_c \approx (4.90\pm0.09)~\bar{n}_{\mathrm{gal}}^{-0.096\pm0.002}\\
    &R_{c,\rm BOSS}^* = 10.52 \pm 0.26 \hmpc,
\end{split}
\end{equation}
with a much better fit for all densities. 

While it is more significant than in the small sphere auto correlation, the dependence on the galaxy density remains small with $\nu = -0.096\pm0.002$. Because of this, we will ignore too this correlation for the rest of the analysis.

\subsubsection{Cross correlation between galaxies and large spheres}
The top-right panel of \reffig{fig:snr-all-r-n-pre} shows the $\SNR$ map obtained from the cross correlations between galaxies and large spheres. We observe a the maximum $\abs{\SNR}$ of $\sim 12$, which is notably higher than that of the galaxy and large sphere auto correlations. Similar to the case of small spheres, the BAO significance of the cross correlation of large spheres and galaxies is higher than that of the large sphere auto correlation.
Besides, the optimised $\SNR$ do not follow a power law in the low density region, where the uncertainties are larger. Our power law fit for this correlation is
\begin{equation}
\begin{split}
    &R_c^* \approx (4.88\pm0.08)~\bar{n}_{\mathrm{gal}}^{-0.138\pm0.002}\\
    &R_{c,\rm BOSS}^* = 14.63 \pm 0.33 \hmpc.    
\end{split}
\end{equation}

For the remainder of the paper, we will consider only the optimal radius cut given by the fit to the auto correlations (\refeq{eq:pre-rdependence}). This is given first, to the fact that the fit to the auto correlation's optima is relatively poor. Second, on larger densities the two fits are very close to each other. Given that the SNR is not a perfect indicator of the BAO measurement precision (see Appendix~\ref{sec:snr-def}), it is reasonable to use the same scaling relations for both the auto and cross correlations. In what follows we will also refer to the sample of large DT spheres as voids.

\subsection{Post-reconstruction}
\begin{figure}
    \centering
    \includegraphics[width=\columnwidth]{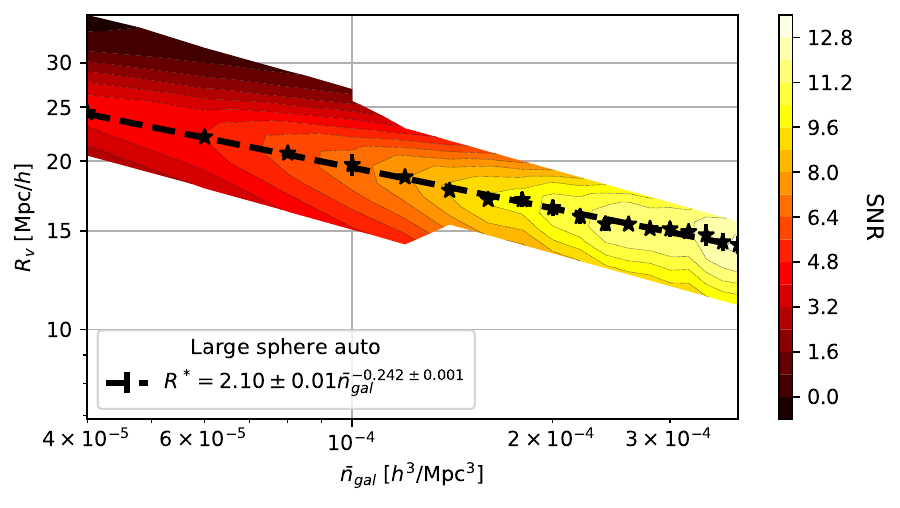}
    \caption{$\SNR(R_c, \bar{n}_{\mathrm{gal}})$ contour map for reconstructed uniformly downsampled boxes. Stars show the $R_c^*$ measurement for each density bin. The dotted line corresponds to the fit to the power-law model.}
    \label{fig:snr-r-n-post}
\end{figure}
We now use the algorithm described in \refsec{sec:gen-methods} to reconstruct the redshift-space \patchy{}-U data. From the large-sphere auto correlations of this data set we obtain the map shown in \reffig{fig:snr-r-n-post}, which yields the relation
\begin{equation}
\begin{split}
    &R_c^* \approx (2.10\pm0.01) ~\bar{n}_{\mathrm{gal}}^{-0.242\pm0.001}\\
    &R_{c,\rm BOSS}^* = 14.41 \pm 0.13 \hmpc.    
\end{split}
    \label{eq:post-rdependence}
\end{equation}
We observe that the $\gamma$ parameter is significantly different from the pre-reconstruction counterpart (\refeq{eq:pre-rdependence}). The smaller $\gamma$ parameter can be explained by the fact that reconstruction narrows the void size distributions, making large voids smaller (see \reffig{fig:void-density-vs-radius}). For instance, for the BOSS density quoted, the threshold before reconstruction is $\sim 15.8 \hmpc$ while after reconstruction it is $\sim14.4\hmpc$.

In contrast, the power parameter $\nu\approx-0.24$ is similar to the one obtained from pre-reconstruction voids. This suggests that the dependence of the cut in the galaxy density is similar in both cases.

\subsection{Summary}
We have found that the SNR-maximising void radius cut depends on the galaxy density through a power law and fit the parameters for different correlations. We focus on the parameters extracted from void auto correlations, which show different amplitudes but similar exponents.

\citet{zhao2017} has shown that the DT spheres with a radius $R_1 = \bar{n}_{\mathrm{gal}}^{-1/3}$ have a bias of 1 and that this relation does not evolve significantly with redshift. Moreover, the same work shows that spheres with $R=R_0\approx1.8\bar{n}_{\mathrm{gal}}^{-0.29}$ have a bias of 0. In this sense, sphere populations with a given bias are related to the galaxy density via power laws. Therefore, we could interpret our power parameter $\nu\approx-0.24$ in a similar manner, as defining a sphere population of negative bias. The redshift evolution of $\nu$ is left for a future work.

\section{Clustering of Voids with non-uniform incompleteness}
\label{sec:clustering-incompleteness}
In this section we first compare the clustering measurements of the \patchy{}-A(R) sets with the \patchy{}-F sample and analyse how the simulated incompleteness affects the correlation functions. We then propose an alternative void sample definition that mitigates some of the discrepancies. Finally, we explore a more realistic incompleteness in light cone mocks, and compare the standard void sample with the one we propose.
\subsection{Simulation boxes}
\subsubsection{The void sample with a constant cut}
\begin{figure}
    \centering
    \includegraphics[width=\linewidth]{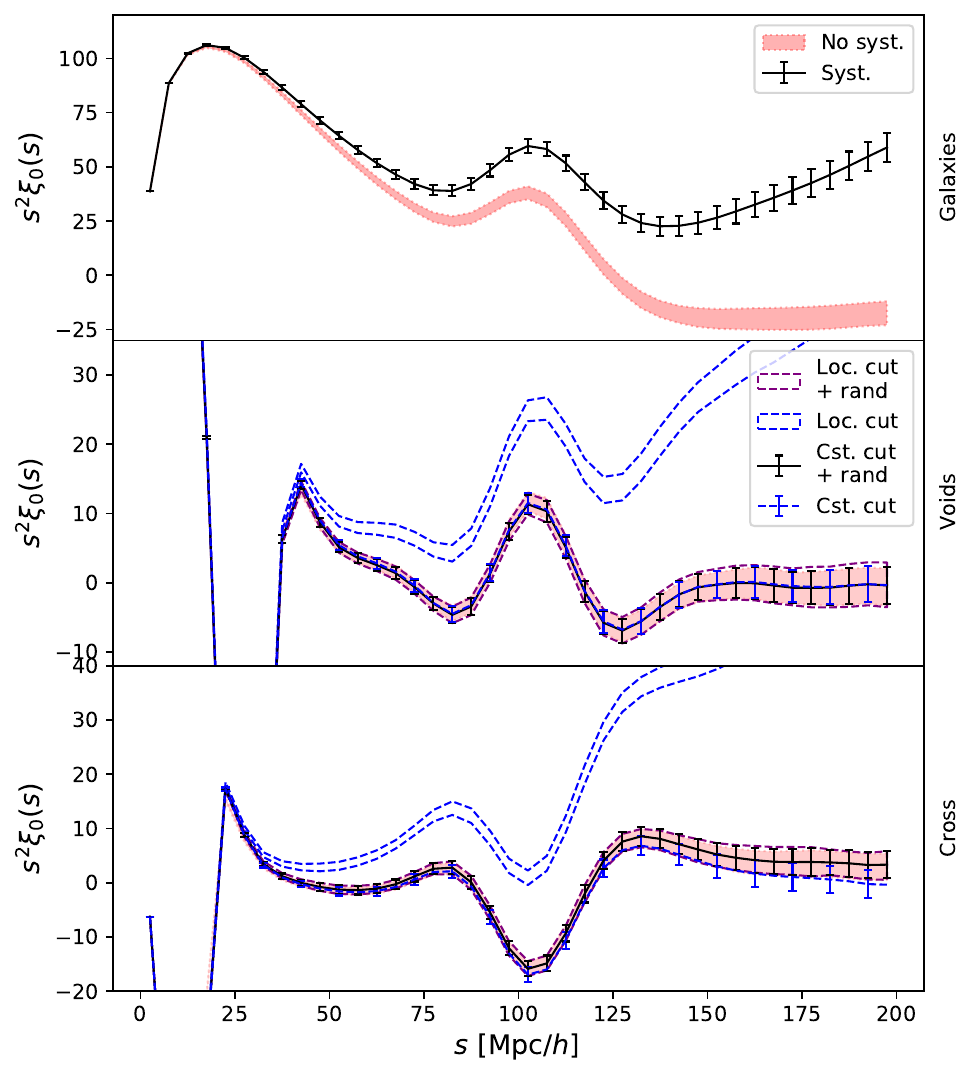}
    \caption{Two-point correlation functions of large-spheres from artificially downsampled boxes that simulate angular systematics. From top to bottom the panels show galaxy-galaxy, void-void and galaxy-void correlations. We show the galaxy-galaxy correlations with (error bars) and without (red envelope) downsampling. For void-void and galaxy-void correlations, we label the use of the local cut with `Loc. cut', contrary to `Cst. cut' that marks the use of a constant cut. When we use randoms taking into account data incompleteness we mark the curve `rand'. Otherwise, analytically computed random pair counts (uniform random sample, no incompleteness) were used.}
    \label{fig:tpcf-angular-avg-loc}
\end{figure}

\begin{figure}
    \centering
    \includegraphics[width=\linewidth]{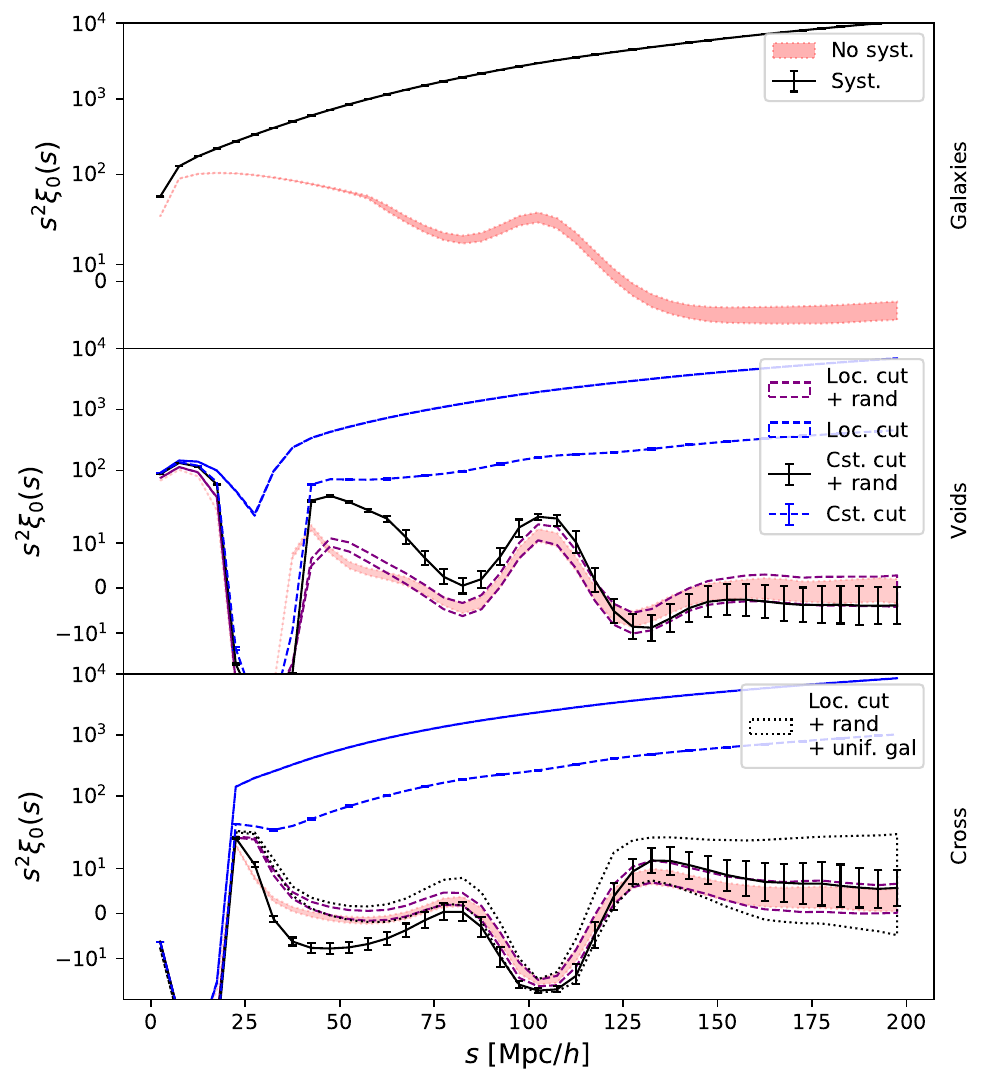}
    \caption{Two-point correlation functions of large-spheres from artificially downsampled boxes simulating radial selection. The color/line convention is the same as in \reffig{fig:tpcf-angular-avg-loc}. The bottom panel shows additionally the galaxy-void correlation where uniform galaxy randoms and non-uniform void randoms are used (dotted envelope).}
    \label{fig:tpcf-radial-avg-loc}
\end{figure}

We show the clustering results in Figs. \ref{fig:tpcf-angular-avg-loc} and \ref{fig:tpcf-radial-avg-loc}. The clustering measurements from \patchy{}-F boxes (marked No syst.), among which the ones involving voids are obtained with a radius cut of $R_c = 15.6\hmpc$, are shown as a red shaded area for reference. For simulation boxes in particular, no weight is used in order to highlight the effects of sample incompleteness on the clustering.

The constant cut void sample is defined as the set
$\mathcal{V}_0 \equiv \qty{\mathtt{s}\in\mathcal{S} | R(\mathtt{s}) > R_c}$, where $\mathcal{S}$ denotes the complete set of DT spheres and $R(\mathtt{s})$ is the radius of the sphere $\mathtt{s}$. In this sample, there is a unique and constant radius cut $R_c$. This is the approach that has been used in previous studies on DT voids \citep{Zhao2016, Liang2016, Kitaura2016, Zhao2020}. In this work, we define $R_c = \gamma\bar{n}_{\mathrm{gal}}^{\nu}$ using the parameters shown in \refeq{eq:pre-rdependence}, which is equivalent to $R_c  = 15.5\hmpc$ for \patchy{}-A and $R_c  = 17.5\hmpc$ for \patchy{}-R. The application of different radius cuts on different data sets is motivated by the strong dependence of the void radius on the galaxy density, as shown in \reffig{fig:void-density-vs-radius}. For instance, the galaxy density of \patchy{}-R is about half that of \patchy{}-F, which means that using the same cut as \patchy{}-F would likely include more small spheres, thus contaminating the void BAO signal.

Given the small difference of galaxy density between the \patchy{}-A and \patchy{}-F samples, the \patchy{}-A correlations (\reffig{fig:tpcf-angular-avg-loc}) show that a small incompleteness is enough to introduce a noticeable bias in the galaxy auto correlation (top panel). This bias is due to the fact that we do not consider the incompleteness of the sample in the random catalogues. This is to highlight the effect of unknown systematics on the samples. Under these same conditions, the void auto correlations remain unscathed, which shows that this void sample is more robust to a moderate incompleteness of the galaxy population than galaxies themselves. 

If we insist in using a uniform random field for estimating the 2PCF of samples with large incompleteness (\reffig{fig:tpcf-radial-avg-loc}), apart of the overestimation of galaxy auto and galaxy-void correlation on large scales, we observe that the bias on the void auto correlation is considerably larger as well. Therefore, for high incompleteness it is mandatory to use appropriate randoms in the estimation of all correlation functions. 

By using void and galaxy randoms that appropriately take into account the large incompleteness of \patchy{}-R data, we obtain correlation functions that closely match \patchy{}-F's on large scales. On lower scales ($\sim 50\hmpc$) however, the void correlations show a significant difference from the \patchy{}-F boxes (\reffig{fig:tpcf-radial-avg-loc}). The 2PCF on this scale is strongly affected by the void exclusion effect, which is related to the maximum size of the voids in the sample \citep{Zhao2016}. The constant cut sample will inevitably include larger DT spheres found in the lower observed density regions of the \patchy{}-R boxes, and enhance the exclusion effect. It should be noted that a similar behaviour is expected of voids obtained from any sparse tracer sample. In fact, strong exclusion features are found in low-density samples such as eBOSS LRG \citep{Zhao2021} and QSO (\textcolor{blue}{Tamone et al. in preparation}).

The robustness of the $\mathcal{V}_0$ sample may be explained by noting that a constant cut of $R_c = 15.5\hmpc$ ($n_{\rm gal}^{1/3}R \in (0.8, 1.2)$) selects spheres that populate the high-density regions of the domain (top-right panel, \reffig{fig:box-incompleteness}). These spheres will fill in the gaps in the field left by the larger spheres ($n_{\rm gal}^{1/3}R \in (1.5, 2.5)$, same figure), which yields a rather uniform void field. Therefore, when estimating the $\mathcal{V}_0$ 2PCF it is enough to consider a uniform random field for normalisation. We must stress that this argument holds for galaxy samples with moderate incompleteness (<20~{\rm per cent}), such as the \patchy{}-A galaxies. When incompleteness is too large (>50~{\rm per cent}), the resulting void sample is not sufficiently uniform.

\subsubsection{The void sample with a local cut}
In Eqs. \eqref{eq:pre-rdependence} and  \eqref{eq:post-rdependence} we show the dependence of the optimal void radius cut on the average densities of uniform galaxy samples in cubic boxes. When working with non-uniform tracer distributions, it is natural to consider a variable cut that uses information about the local galaxy density instead of its average over the domain. Consequently, we define a new void sample $\mathcal{V}_l \equiv \qty{\mathtt{s}\in\mathcal{S} | R(\mathtt{s}) > \gamma n_{\mathrm{gal}}(\vb*{x}_{\mathtt{s}})^\nu}$, where $\vb*{x}_{\mathtt{s}}$ denotes the centre of the sphere $\mathtt{s}$. This sample will be often referred to as the local cut sample given that it depends on the DT sphere position. The 2PCFs involving these void samples are shown as a dashed-line envelope in Figs. \ref{fig:tpcf-angular-avg-loc} and \ref{fig:tpcf-radial-avg-loc}. 

We find that the correlation functions of the void samples with local radius cuts are biased towards large scales if the randoms do not account for the incompleteness, even when it is small, as evidenced by the blue envelopes in \reffig{fig:tpcf-angular-avg-loc}. This behaviour is expected, since the final void distribution is -- by construction -- closely related to the incompleteness and cannot therefore be well approximated by a uniform field.  On the contrary, no bias is observed when using the void randoms generated using the shuffling algorithm in \refsec{sec:2pcf-estimation}. These results indicate that the $\mathcal{V}_l$ sample is more sensitive to unknown systematics than $\mathcal{V}_0$. Thus we must always use appropriate void randoms.

Moreover, \reffig{fig:tpcf-radial-avg-loc} also shows that the void exclusion feature obtained with the local cut (error bars) is less prominent than the one obtained using the constant cut (purple dashed envelope) when there is a large incompleteness in the data. In fact, the $\mathcal{V}_l$ sample yields clustering measurements that are consistent with those of the \patchy{}-F data on all scales in the presence of severe incompleteness. The reason for the discrepancy in the exclusion patterns of both void samples can be explained by the presence of largely undersampled regions in the data. Consider, for example, the $Z\sim0$ and $Z\sim2500\hmpc$ edges of the \patchy{}-R box, where the galaxy number density is the smallest, so the local average void radius is  $~23\hmpc$. An object selection based on the constant cut obtained from the average density over the whole box ($R_c = 17.5\hmpc$) means that a large number of average-sized DT spheres in the region is selected, even though they may not trace matter underdensities. That being the case, these spheres contribute to the higher amplitude of the void exclusion pattern (\reffig{fig:tpcf-radial-avg-loc}). Conversely, a selection performed with the local cut of $R_c\approx25\hmpc$ successfully discards average-sized spheres. As a result, smaller DT spheres that `pollute' the sample are ignored, and the exclusion effect becomes less significant. 

Finally, we use the \patchy{}-R data to investigate the robustness of the local cut to vastly inadequate randoms for galaxies (see top panel in \reffig{fig:tpcf-radial-avg-loc}); given that in practice, the void randoms for the observational data are always generated using the shuffling method \citep{Liang2016}. To this end, we compute the 2PCFs using uniform randoms for galaxies. In this case, despite the larger variance, the correlation function shows little bias, even using the local cut. This indicates that even if we are unable to correct for unknown systematics in the galaxy data, the cross correlation can still be unbiased using the local cut and the void randoms obtained through shuffling. The robustness of the cross correlation and the fact that it has a larger $\abs{\SNR}$ than both galaxy and void auto correlations make it a promising tool for BAO analysis beyond current standard techniques.

In summary, we find that the constant cut is very robust to low sample incompleteness, such as the ones caused by angular systematic effects of a galaxy survey, as regions with low angular completeness are typically masked in the galaxy catalogues for clustering analysis \citep{Ross2011, Ross2017a}. In this case we are able to recover the correlations that are expected from a complete sample without observational systematics. On the contrary, when the incompleteness is very high, the void sample suffers from the same systematics as galaxies if the geometry of the survey is not well accounted for in the randoms. Furthermore, even using appropriate (shuffled) randoms, the void-void correlation using the $\mathcal{V}_0$ sample shows a large exclusion peak due to the inclusion of very large spheres from undersampled regions. In this high incompleteness case, the local cut proposed here avoids these DT spheres and yields a correlation that is compatible with the complete-box measurements on all scales. We have also shown that the cross correlation is robust even to very large incompleteness in the galaxy sample as long as appropriate void randoms (shuffled) are used.

\subsection{Light cones}
\begin{figure*}
    \centering
    \includegraphics[width=\linewidth]{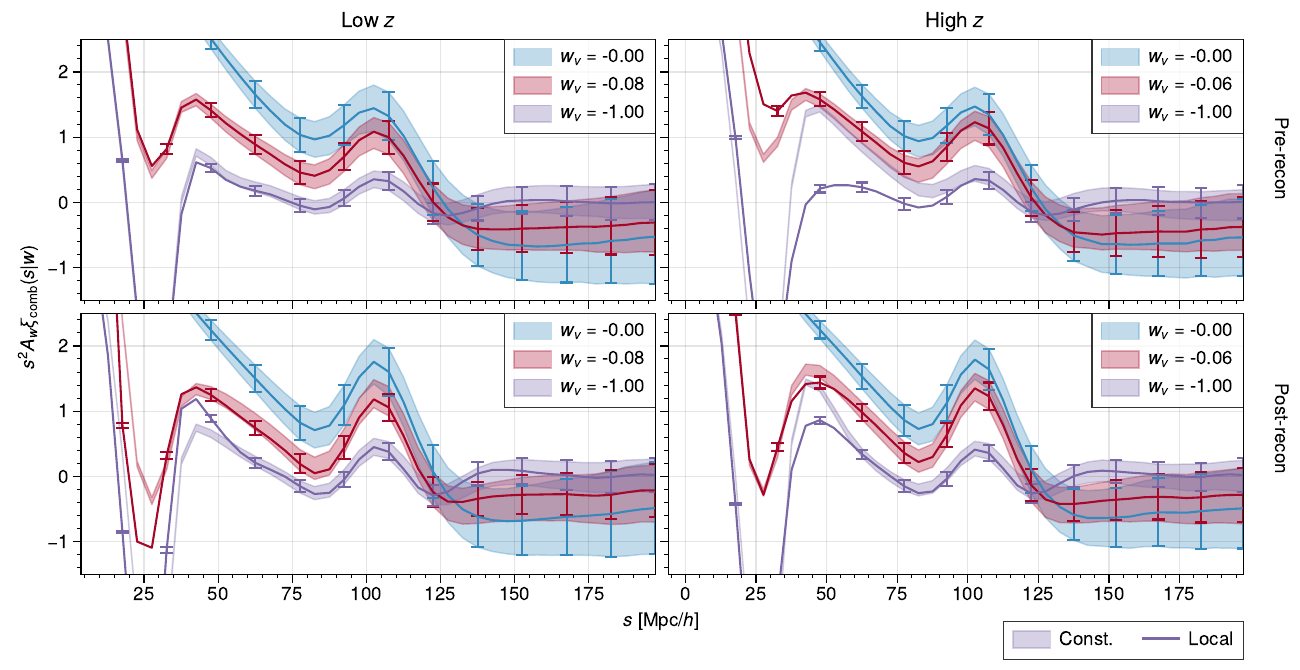}
    \caption{Average normalised combined two-point correlation functions from mock light cone data using different $w_v$ (see \refsec{sec:bao-constraints} for the choice of $w_v$). The normalisation is defined as $A_{w_v}^{-1} = \int_0^{200}\xi_{\rm comb}(s|w_v)~\dd s$.  
The shaded envelopes show the clustering obtained using the void sample extracted with a constant cut ($\mathcal{V}_0$). In this case we have chosen $R_c=16\hmpc$ to contrast with the results of \citet{Zhao2020}. The error bars show the clustering results using a local cut ($\mathcal{V}_l$) to define the void sample. We show the low-$z$ redshift bin on the top row and high-$z$ on the bottom row. The top row displays pre-reconstruction clustering while the bottom shows post-reconstruction measurements.}
    \label{fig:void-tpcf-avg}
\end{figure*}
In light cone data, the largest source of incompleteness is typically the radial selection function. In what follows, we will assume that angular incompleteness is sufficiently small compared to the radial incompleteness. Thus $n_{\mathrm{gal}}(\vb*{x}) \approx n_{\mathrm{gal}}(z)$. The radial selection function depicted in \reffig{fig:nz-caps} shows that the number densities of LRGs are more stable than those of the high-$z$ bin, in which density drops significantly as $z$ approaches $0.75$. We define the constant cut void sample with a radius of $R_c=16\hmpc$ in order to be consistent with the results in \citet{Zhao2020} for comparison purposes.
 
 We observe that for pre-reconstruction light cones (top row in \reffig{fig:void-tpcf-avg}) the low-$z$ clustering (left column) is consistent for both the constant cut void sample (shaded envelope) and the local cut sample (error bars). The absence of significant differences in the auto correlations can be attributed to the fact that the galaxy density does not vary significantly in this redshift range (see \reffig{fig:nz-caps}). In this case the local cut is near constant. In contrast, the high-$z$ constant cut clustering in the right panels of \reffig{fig:void-tpcf-avg} shows the same low-scale feature observed in the \patchy{}-R 2PCF, which can be explained by the large drop in galaxy density at high redshift. The use of a local cut (\refeq{eq:pre-rdependence}) succeeds in better selecting a void sample, discarding the large spheres that reside in low sampling regions that enhance the exclusion effect.
 
 When we instead use reconstructed light cones and the model in \refeq{eq:post-rdependence} we obtain results that mirror the behaviour of pre-reconstruction clustering. Additionally, it is clear that the void exclusion in the high-$z$ 2PCF (bottom-right panel) using the constant cut is not as wide as without reconstruction, and is therefore less intrusive at BAO scales. This alone already enhances the BAO peak SNR. A narrower exclusion peak can be explained by a smaller number of the largest voids in the reconstructed catalogue. In fact, the void size distributions of reconstructed cubic mocks show that there is a smaller fraction of large DT spheres (see \reffig{fig:void-density-vs-radius}, bottom panel). An analogous behaviour of the void distributions of the light cones would explain the narrower exclusion peak.
 
\begin{figure}
    \centering
    \includegraphics[width=\columnwidth]{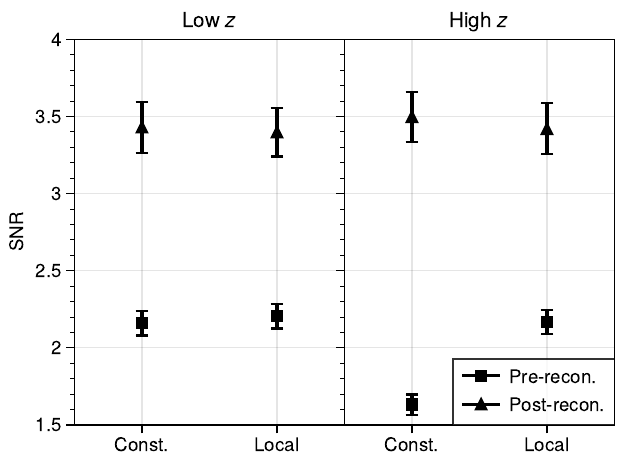}
    \caption{$\SNR$ computed from MD-\textsc{Patchy} DR12 light cone voids using both the constant and the local radius cut  (Eqs. \ref{eq:pre-rdependence}, \ref{eq:post-rdependence}). Squares denote the post-reconstruction results while triangles show pre-reconstruction. The error bars correspond to the jackknife 95~{\rm per cent} confidence interval}
    \label{fig:snr-voids}
\end{figure}
 The BAO $\SNR$ measured from pre- and post-reconstruction void-void correlations in different redshift bins are shown in \reffig{fig:snr-voids}. The local cut yields significantly higher $\SNR$ only in the high $z$ bin before reconstruction. For all other cases, the SNR is consistent for both local and constant cuts. After reconstruction especially, the improvement is not particularly significant, given that the displacement already suppresses large-scale exclusion effects. This shows that despite its simplicity, the constant cut is a near-optimal void selection criterion that maximises the significance of BAO extracted from voids that are constructed upon reconstructed galaxy samples.

 \section{BAO constraints with incomplete samples}
\label{sec:bao-constraints}
\begin{figure}
    \centering
    \includegraphics[width=\linewidth]{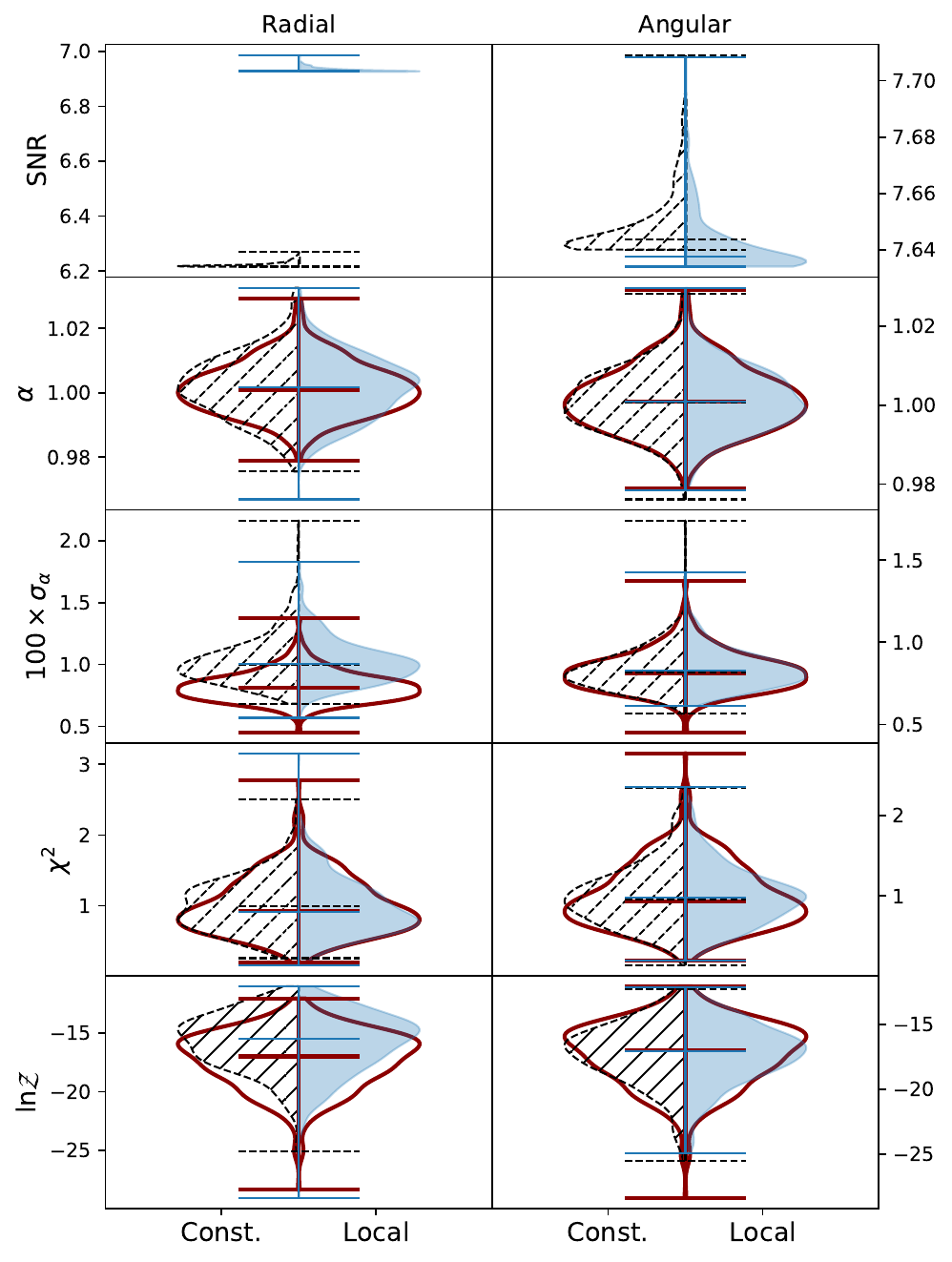}
    \caption{Void-only BAO fit results and SNR measurements. Top row shows the SNR measurements on \patchy{}-R (left) and \patchy{}-A (right) data. Panels below show BAO statistics such as the $\alpha$ parameter measurement, the error on its estimation $\sigma_\alpha$, the goodness of fit $\chi^2/\mathrm{dof}$ with $\mathrm{dof}=12$ (number of degrees of freedom in the fit) and the Bayesian evidence expressed as $\ln{\mathcal{Z}}$. Distributions are obtained from fitting each individual realisation in the data sets. The left half of the violin plots (hatched, dashed line) represents the distribution of the corresponding statistic obtained using a constant cut. The shaded distributions (continuous line) on the right shows the analogous results using a local cut. The violin with a continuous line (red) shows the BAO fit to complete boxes. The top, middle and bottom horizontal lines on the violins are the maximum, median and minimum of each distribution, respectively.}
    \label{fig:baostats-nu}
\end{figure}

Our analyses so far are all based on the SNR estimation given by \refeq{eq:snr-estimate}. In order to quantify the impacts of sample incompleteness on BAO peak position measurements, we first fit the void auto correlations from the downsampled boxes using the parabolic model introduced in \refsec{sec:gen-methods}. We fit the correlations computed taking into account the non-uniformity of the samples in order to have unbiased 2PCFs in all cases. \reffig{fig:baostats-nu} shows the SNR measurements and BAO fit results. We observe in the top panel that the use of a local cut is advantageous in terms of $\SNR$ in the large incompleteness (radial selection) situation. Meanwhile the difference in SNR for small incompleteness (right hand side of \reffig{fig:baostats-nu}) is seen to be insignificant, and no clear advantage of using a local cut is observed.

On the same figure, the second row shows that the $\alpha$ parameter is recovered in all cases with negligible differences with respect to the \patchy{}-F measurements (shown as red curves). The distribution of the error on the BAO scale $\sigma_\alpha$ is shown to be consistent among the downsampled boxes and the complete \patchy{}-F sample. Nonetheless, when incompleteness is large, the distribution of the error shows a slightly positively biased mean with respect to \patchy{}-F. This small shift is explained by the fact that large incompleteness causes a larger exclusion effect. The goodness-of-fit metrics $\chi^2$ and $\ln{\mathcal{Z}}$ of the downsampled boxes do not show significant differences when compared to the complete sample either. 

In summary, the BAO fit results show that the local and constant cuts provide compatible results in terms of $\alpha$ parameter and error on this estimate. Additionally, both the constant and local cuts applied to the downsampled data provide results unbiased from those coming from the complete \patchy{}-F sample.

When using light cones, the data volume is reduced and the fits to void auto correlations are more difficult. To improve the significance of the signal for the fit we use instead the combined 2PCF. This implies that we must first optimise the clustering measurements by tuning the $w_v$ parameter of the modified LS estimator. Thus, we explore various $w$ values and perform the BAO fit on the average of the mocks as well as evaluating the BAO $\SNR$ for each $w_v$. We choose values of $w_v$ in the range of $(-0.2, -0.01)$. These values are negative to account for the fact that voids and galaxies are anti-correlated, which means that the peak in the cross-correlation is negative. Thus, $w_v<0$ is required in order to add the contribution of the cross correlation.

\begin{figure}
    \centering
    \includegraphics[width=\columnwidth]{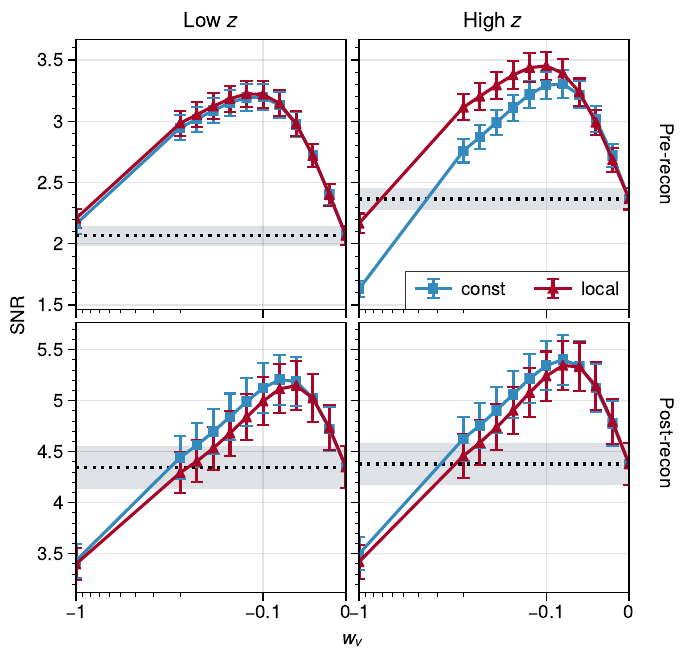}
    \caption{$\SNR$ as a function of $w_v$ computed from light cone $\xi_{\mathrm{comb}}$. Left and right columns show the low and high $z$ bins respectively. Top row corresponds to pre-reconstruction and bottom row to post-reconstruction results. The error bars correspond to the jackknife 95~{\rm per cent} CI. The dotted lines correspond to the estimated galaxy $\SNR$ and the shaded areas denote the 95 {per cent} CI. $\SNR$ values over this region show improvement due to the inclusion of void information.}
    \label{fig:snr-comb}
\end{figure}
\begin{figure}
    \includegraphics[width=\linewidth]{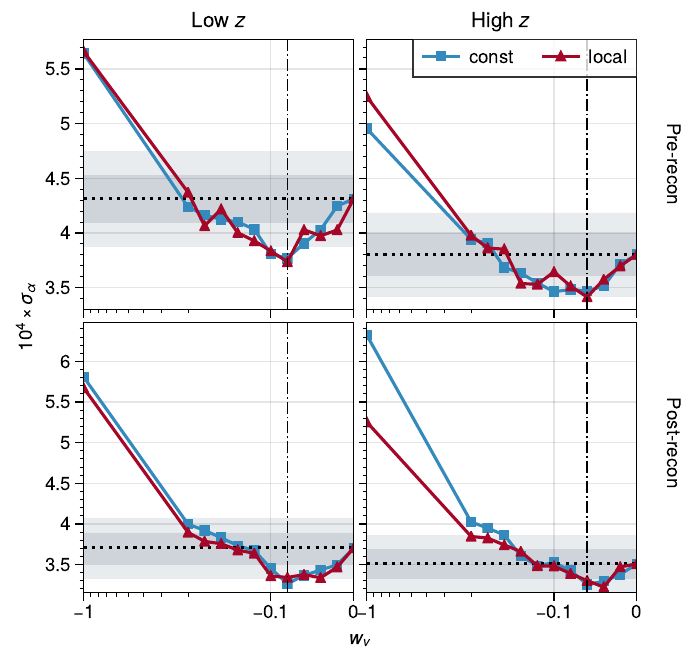}
    \caption{$\sigma_\alpha$ obtained from the fit to the mean of 2048 (1000) pre- (post-) reconstruction light cone $\xi_{\mathrm{comb}}$ as a function of $w_v$. Top row displays pre-reconstruction and bottom row shows post-reconstruction results. The low $z$ results are shown on the left and the high $z$ on the right. The vertical dotted lines correspond to the chosen values of $w_v$ for each sample. The horizontal dotted lines show the galaxy-only ($w_v=0$) error on alpha. Values below this line indicate improvement from using void information. The shaded areas correspond to 5 and 10 per\,cent bands around this value.}
    \label{fig:bao-avg-comb}
\end{figure}

The top panels in \reffig{fig:snr-comb} shows the $\SNR$ vs. $w_v$ relations with both the local and constant cuts for catalogues before reconstruction. In the low $z$ bin (left), the relationships for the constant and local cuts are very similar. This is expected given the similarities between the void correlations for this bin (\reffig{fig:void-tpcf-avg}). In the high $z$ bin (right), however, there is a significant $~95~{\rm per cent}$ confidence interval (CI) increase in the $\SNR$ of the local cut when compared to the constant cut. This is also compatible with the results from the void $\SNR$ only. The bottom panels of \reffig{fig:snr-comb} display the corresponding post-reconstruction results. Evidently, the difference between the void samples is not significant according to the displayed jackknife confidence interval. As a reference we include the $\SNR$ value of a galaxy-galaxy correlation along with its 95 {\rm per cent} CI. The SNR clearly improves when using void information with $w_v\sim-0.1$ but drops as the void terms start dominating the combination.

\reffig{fig:bao-avg-comb} shows the $\sigma_\alpha$ distribution of the BAO fit performed with the different $w_v$ parameters. Analogously to the procedure used when fitting individual mocks, we first fit the average correlation with a free $c$ parameter and then redo the fit with $c$ fixed to the best-fit estimation. Before reconstruction $\sigma_\alpha$ is not significantly sensitive to the void sample selection scheme and suggests an optimal value of $w_v\approx-0.08$ for the low $z$ bin and $w_v\approx-0.06$ for the high $z$. Notice that when $w_v$ decreases, the error on $\alpha$ increases significantly as the contribution of voids becomes more significant; when $w_v=0$ the results reduce to constraints from galaxies alone, represented as horizontal lines in \reffig{fig:bao-avg-comb}. The same figure shows that an average improvement of 10 per~cent in $\sigma_\alpha$ can be obtained by including voids with a given $w_v$. After reconstruction we find that the same $w_v$ values provide a 5 to 10 per~cent improvement in the BAO scale measurement depending on the redshift bin.

Comparing the plots for SNR (\reffig{fig:snr-comb}) and BAO fits (\reffig{fig:bao-avg-comb}) we observe that the relationships of both the SNR and the error on alpha with $w$ are consistent with each other as long as galaxies are the dominating terms in the combination. This may be due to the fact that our current BAO model does not capture the void exclusion perfectly, thus reducing the reliability of the fits as the void terms become dominant. The samples with larger SNR show consistently smaller error on $\alpha$. In particular, the pre-reconstruction high $z$ sample (top-right panel of Figs. \ref{fig:snr-comb} and \ref{fig:bao-avg-comb}) shows the expected gap in the performance between the local and constant cuts, where the local selection yields higher SNR and lower $\sigma_\alpha$. We observe that even though this gap is significant over 95~per~cent in the SNR, the difference in $\sigma_\alpha$ is much smaller and not considered significant. The agreement in the behaviour of the SNR and the $\sigma_\alpha$ suggests that a large significance of the BAO peak is a good indicator of a low $\sigma_\alpha$.

In summary, for pre- and post-reconstruction we use the values of $w_v=-0.08$ in the low and $w_v=-0.06$ for the high $z$ bins with both constant and local cuts. Even though these values are small in magnitude, the precision of BAO measurement from the corresponding combined correlation functions is improved  by 5 to 10 {\rm per\,cent} -- with respect to the measurements obtained from galaxy correlations alone ($w_v=0$) -- when including information coming from the void samples. The combined correlations given the chosen $w_v$ values are shown in \reffig{fig:void-tpcf-avg}. The magnitude of the void weight $w_v$ implies that a small amount of void information is used in order to improve the BAO measurement. Indeed, \citet{Zhao2021} have shown that it can lead to an improvement in cosmological parameter estimation comparable to standard anisotropic BAO analyses.

\begin{figure}
    \centering
    \includegraphics[width=0.99\linewidth]{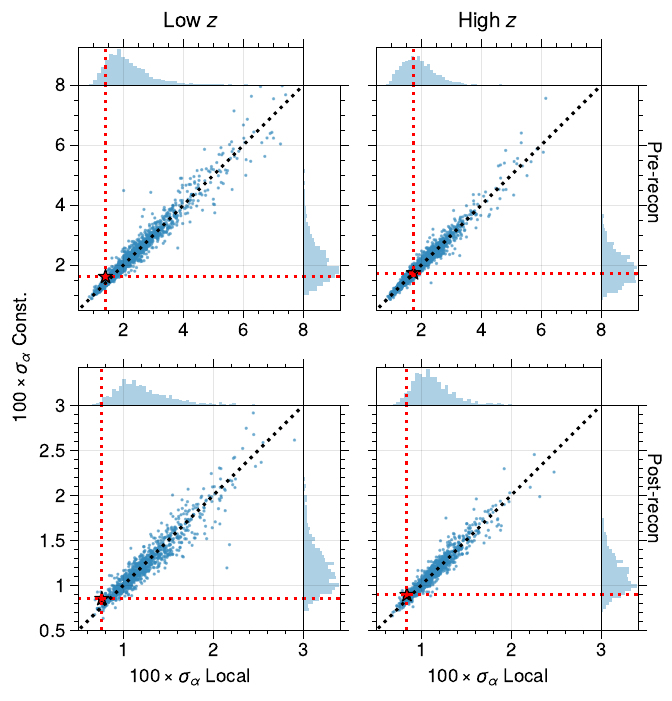}
    \caption{Distribution of $\sigma_\alpha$ fit from MD-\patchy{} light cones and BOSS DR12 data. We compare the constant and the local cuts. \textit{Top} Pre-reconstruction. \textit{Bottom:} After reconstruction. \textit{Left: } Low $z$ redshift bin ($z\in(0.2, 0.5)$). \textit{Right: } High $z$ redshift bin  ($z\in(0.5, 0.75)$). The stars and dotted vertical and horizontal lines show the error on alpha obtained from the fit to BOSS DR12 data. The diagonal dotted line represents equality.}
    \label{fig:bothcut-comparison}
\end{figure}

Using these $w_v$ values, we perform individual fits on each mock to analyse the distribution of $\sigma_\alpha$, these results are shown in \reffig{fig:bothcut-comparison}. We have also fixed $c$ to the best-fit value from the fit to the average of the mocks. We compare the constant cut ($y$-axis) and the local cut ($x$-axis) measurements. \patchy{} BOSS DR12 light cones are shown as dots and BOSS DR12 observations as stars. The top panels display the results from pre-reconstructed data while the bottom panels show analogous measurements on reconstructed light cones. The results for the constant cut here shown are consistent with those in \citet{Zhao2020}. Our fits using the local cuts show no significant difference in terms of BAO measurements compared to the constant cut. Fits to the BOSS observations show errors consistent with the cosmic variance of the individual mocks, which implies there is no clear advantage of one void sample selection scheme over the other.

\section{Conclusions}
\label{sec:conclusions}
Throughout this work we have analysed the response of DT void sample and its clustering to systematic incompleteness in the tracer distribution. For this purpose we have prepared the Non-Uniformly Angular downsampled (\patchy{}-A) and Non-Uniformly Radial downsampled (\patchy{}-R) data sets. They are constructed by downsampling the full \patchy{} boxes with a paraboloid in the $X,Y$ plane and with a Gaussian selection function in the $Z$ direction respectively. With this samples we aim at modelling angular systematics and radial selection effects of galaxy surveys. In addition, we have studied the dependence of the DT void radii distribution with the tracer density before and after the reconstruction process.

We find fitting formulae to describe the void sample selection criteria as a density-dependent threshold radius $R_c(n_{\mathrm{gal}};\gamma,\nu)=\gamma n_{\mathrm{gal}}^\nu$. We studied auto and cross correlations with either small or large spheres. The BAO significance from the auto correlations of large spheres -- under-dense regions that are typically referred to as cosmic voids -- is found to be sensitive to the tracer density. The fit parameter $\nu$ is similar in pre- and post- reconstruction fits, while $\gamma$ varies. This suggests that to obtain the optimal radius cuts for different galaxy densities, one can fix the $\nu$ parameter and grid-search for an appropriate $\SNR$-maximising $\gamma$. Moreover, we observe that the cross correlation between galaxies and large voids has a larger $\abs{\SNR}$ than even the galaxy-galaxy correlation, which is the current standard for BAO measurements. 

In terms of the broad-band clustering, we show that the large-void sample defined as $\mathcal{V}_0 \equiv \qty{\mathtt{s}\in\mathcal{S} | R(\mathtt{s}) > R_c}$ (constant cut), is not sensitive to systematic effects that manifest as a small sample incompleteness. In contrast, the effects of a very large incompleteness, such as the one generated by the radial selection function, do induce large biases in the galaxy-galaxy, galaxy-void and void-void correlations. This indicates that a more careful estimation of the survey geometry, i.e. generation of random catalogues, is needed in this case. Taking this into account, we find that the $\mathcal{V}_0$ sample shows a large void exclusion pattern at low scales that is more prominent than the expectation from the correlations of uniform \patchy{} data.  An alternative definition of the void sample $\mathcal{V}_l\equiv \qty{\mathtt{s}\in\mathcal{S} | R(\mathtt{s}) > \gamma n_{\mathrm{gal}}(\vb*{x}_{\mathtt{s}})^\nu}$ that takes into account the local tracer density, greatly reduces the amplitude of this feature and yields clustering that is much more consistent with the uniform \patchy{} sample at all scales. The excess probability in the void exclusion pattern can be explained by the fact that the constant radius cut defining the $\mathcal{V}_0$ sample allows for the selection of large spheres in undersampled regions of space that do not necessarily correspond to matter-field underdensities.

We compare the clustering of the $\mathcal{V}_0$ and $\mathcal{V}_l$ samples from \patchy{} light cones pre- and post-reconstruction. Similar to the case of boxes, the pre-reconstruction void-void correlations of the $\mathcal{V}_l$ sample show a significantly lower exclusion in the high-$z$ bin, which yields a more prominent BAO peak. After reconstruction, the exclusion pattern is narrower and it no longer affects BAO scales significantly. For this reason, using the local cut does not seem to affect the SNR significantly in this case.

Furthermore,  we evaluate the impact of the different void selection mechanisms on cosmological measurements. We observe on cubic data as well as on light cones, that the differences in terms of BAO scale measurement precision are negligible and that the $\alpha$ parameter is always recovered. In addition, for the reconstructed samples, the exclusion feature is less intrusive upon BAO scales, even when using the simpler constant cut selection scheme. Therefore, we conclude that for reconstructed samples with BOSS (LRG-like) density, such as \patchy{} and BOSS DR12, the constant cut already delivers a robust clustering sample that allows for an unbiased measurement of the BAO peak position. In particular, this sample is shown to be robust even to unknown systematics that represent a moderate incompleteness (<20~{\rm per cent}), for which the galaxy auto correlation cannot be corrected. Under these conditions, the $\mathcal{V}_0$ sample produces not only an unbiased void auto correlation, but also an unbiased void-galaxy correlation with a generally more significant BAO signal than both auto correlations.

Sparser observations may however be susceptible to large exclusion patterns that can intrude upon BAO scales. In these cases the local radius cut might play a more significant role in the void BAO measurement. Alternatively, the presence of this feature could be modelled and taken into account more precisely in the BAO fits. We leave this analysis for a future work. 

We believe that the approach described here, can in principle provide tighter BAO constraints than just the galaxy auto correlation measurement, and we plan to apply this methodology to the upcoming DESI data.

\section*{Acknowledgements}
DFS, CZ, AV and AT acknowledge support from the SNF grant 200020\_175751. FSK acknowledges the grants SEV-2015-0548, RYC2015-18693, and AYA2017-89891-P.

\section*{Data Availability}

The light cone mocks and BOSS DR12 data are publicly available. The \patchy{} boxes used can be provided upon request to CZ.



\bibliographystyle{mnras}
\bibliography{refs} 




\appendix

\section{Robustness of the BAO SNR definition}
\label{sec:snr-def}
In this section, we justify the choice of the BAO SNR definition given by \refeq{eq:snr-estimate}, which provides an efficient and model-independent way to assess the precision of BAO scale constraint from different 2PCFs. To this end, we compare the SNR estimates and BAO fitting errors for different cases.

Reliable BAO fitting errors require precise BAO models for the clustering of voids with different radii. Unfortunately, the parabolic model introduced in \refsec{sec:baomodel} is not able to model well the void-void and galaxy-void correlations separately {\textcolor{blue}{(Variu et al. in preparation)}}. An alternative to the parabolic term in the void model is to use a template for the function $P_{\rm v, nw}(k)$ obtained from Gaussian random fields. In this approach, we are able to model the void exclusion effect on the 2PCF more accurately for different radius cuts. A complete description of this approach can be found in \citet{Zhao2021} and \textcolor{blue}{Variu et al. (in preparation)}. 

\begin{figure}
    \centering
    \includegraphics[width=\linewidth]{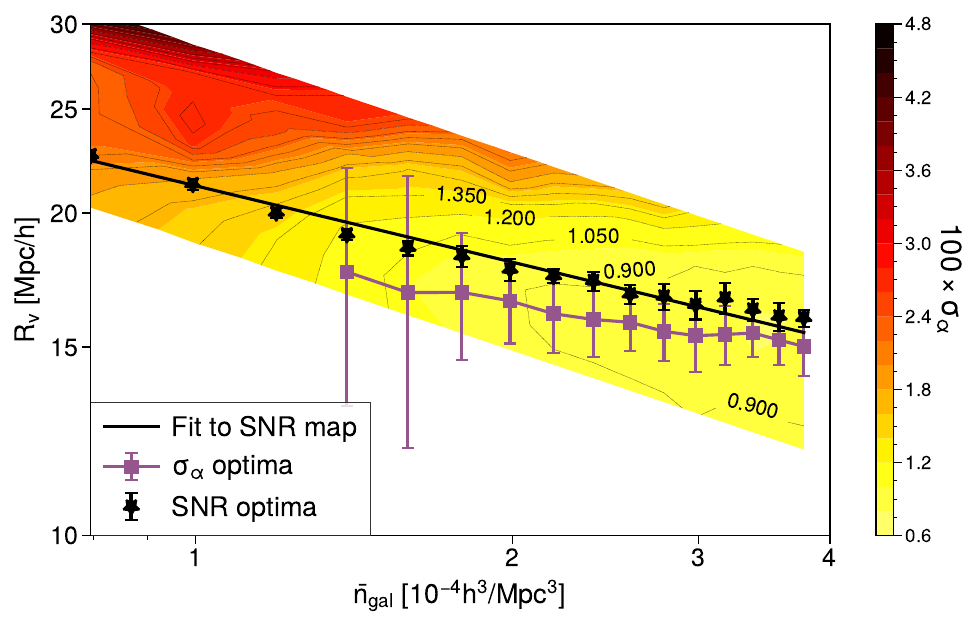}
    \caption{Contour map of $\sigma_\alpha$ as a function of galaxy density and radius cut for pre-reconstruction void-void auto correlations. The $\sigma_\alpha$ are estimated from the median value of the ensemble of measurements from a set of mocks. The standard deviation was estimated as the average of the distances between the 16th and 64th percentiles to the median. The squares (purple) show the estimated optimal radii obtained from minimising $\sigma_\alpha$ along with the estimated error. For reference we show also the optimal radii obtained from the SNR maximisation (stars) and the fit to those points (continuous line, \ref{eq:pre-rdependence}). Contour labels also show values of $100\times\sigma_\alpha$.}
    \label{fig:sigmaa_map}
\end{figure}

We use the template method to obtain a $\sigma_\alpha(R_c, \bar{n}_{\mathrm{gal}})$ map by individually fitting the void-void correlations from the \patchy{}-U sample. \reffig{fig:sigmaa_map} shows the contour map obtained analogous to the top left panel of \reffig{fig:snr-all-r-n-pre}. Contrary to the SNR map, which shows an evident maximum in each density bin, the minimum $\sigma_\alpha$ is relatively insensitive to the choice of void radius cut, as is evident from the wide contours shown in \reffig{fig:sigmaa_map}. For some large-enough densities ($\bar{n}_{\mathrm{gal}} > 1.4\times10^{-4}~h^3\rm Mpc^{-3}$) we are able to find optimal radii using the same technique described in \refsec{sec:gen-methods} but minimising $\sigma_\alpha$ instead of maximising SNR. The optimal radius cuts obtained by minimising the $\sigma_\alpha$ curves are shown as purple squares in \reffig{fig:sigmaa_map}. We do not show densities for which we did not find a minimum within the sampled domain. The the optimal radii obtained from maximising the SNR (black stars) are shown to be consistent with the minima of the $\sigma_\alpha$ to a 1$\sigma$ level. 

\begin{figure}
    \centering
    \includegraphics[width=\linewidth]{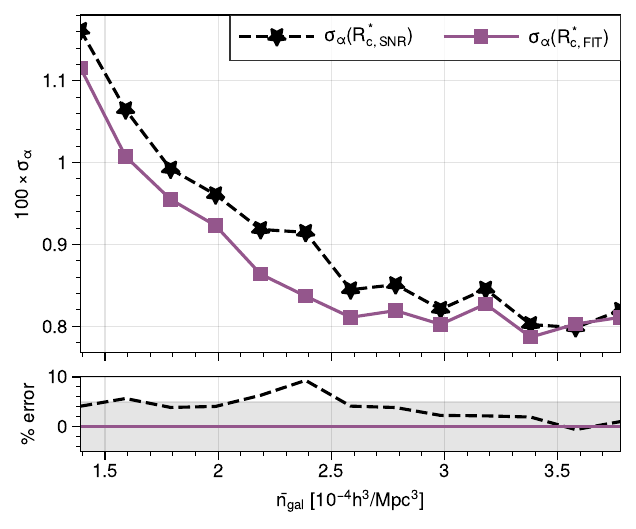}
    \caption{Values of $\sigma_\alpha$ interpolated at the optimum radii obtained using both the SNR maximisation $R^*_{c,\rm SNR}$ and the $\sigma_\alpha$ minimisation $R^*_{c,\rm FIT}$. Values are shown for a subset of the density bins where $R^*_{c,\rm FIT}$ lies within the explored domain. The sub-panel shows the percent error between the different estimations and the shaded area corresponds to the 5~{\rm per cent} error.}
    \label{fig:snr-sigma-calibration}
\end{figure}

In addition, \reffig{fig:snr-sigma-calibration} shows the $\sigma_\alpha$ values interpolated at the optimal radii deduced both from the BAO fit $R^*_{c,\rm FIT}$ and from the SNR maximisation $R^*_{c,\rm SNR}$. Both sets of values are within 5~{\rm per cent} agreement with each other for most density bins, with the SNR-deduced value being systematically larger than the real minima obtained from the BAO fits.

\begin{figure}
    \centering
    \includegraphics[width=\linewidth]{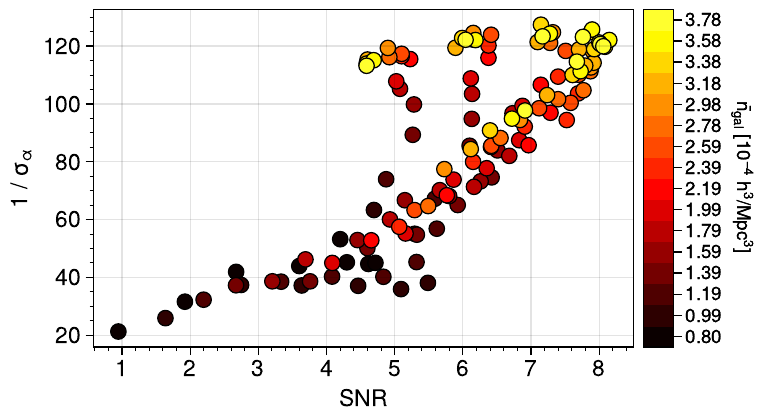}
    \caption{Inverse error on the fitted BAO scale ($1/\sigma_\alpha$) as a function of the BAO SNR as defined in \refsec{sec:2pcf-estimation}. Colorbar shows different galaxy densities.}
    \label{fig:snr-vs-invsigmaa}
\end{figure}

Moreover, \reffig{fig:snr-vs-invsigmaa} shows the relation between the inverse $\sigma_\alpha$ and the $\SNR$ for different galaxy densities. From this plot it is clear that a high SNR is correlated to a small $\sigma_\alpha$. These tests show therefore that the definition of SNR used through the paper is a good proxy for the error in the BAO scale position, $\sigma_\alpha$. 

Finally, we stress that the BAO fit results are model-dependent and sensitive to hyperparameters such as the fitting range. Indeed, the map in \reffig{fig:sigmaa_map} shows less density bins that the analogous SNR map (\reffig{fig:snr-all-r-n-pre}, top-left) because we are not able to obtain good BAO fits for low densities. This is because when density is low, the exclusion pattern affects BAO scales and dominates the fit, such that the fitted parameters are not reliable under such conditions. The SNR provides then a robust, computationally inexpensive alternative for situations where a good BAO model is not yet available.



\bsp	
\label{lastpage}
\end{document}